\documentclass{article}
\usepackage[utf8x]{inputenc}
\usepackage[margin=1in]{geometry}
\usepackage{epsfig} 
\usepackage{ulem}
\usepackage{color, soulutf8}
\usepackage{amsmath}
\usepackage{amssymb}
\usepackage{amsthm}
\usepackage{lscape}
\usepackage{dsfont}
\usepackage{bbm}
\usepackage{url}
\usepackage[colorlinks,bookmarksopen,bookmarksnumbered,citecolor=red,urlcolor=blue]{hyperref}
\usepackage[superscript, biblabel]{cite}
\usepackage{calc}
\usepackage{setspace}
\newcommand{\bX}{ {\boldsymbol{X}} }
\newcommand{\bY} { {\boldsymbol{Y}} }

\newcommand{\bBeta} { \boldsymbol{\beta} }
\newcommand{\bBetatilde} { \boldsymbol{\widetilde{\beta}} }
\newcommand{\bmu} {\boldsymbol{\mu} }
\newcommand{\bmudiff} {\boldsymbol{\mu}^{\textbf{diff}} }
\newcommand{\bSigma} {\boldsymbol{\Sigma}}

\newcommand{\bTheta} { \boldsymbol{\Theta} }
\newcommand{\bT} { \boldsymbol{T} }

\newcommand{\wildetilde} {\widetilde}

\newsavebox\CBox
\newcommand\hcancel[2][0.5pt]{%
  \ifmmode\sbox\CBox{$#2$}\else\sbox\CBox{#2}\fi%
  \makebox[0pt][l]{\usebox\CBox}%
  \rule[0.5\ht\CBox-#1/2]{\wd\CBox}{#1}}

\newtheorem*{theorem*}{Theorem 1}
\newcommand\blfootnote[1]{%
  \begingroup
  \renewcommand\thefootnote{}\footnote{#1}%
  \addtocounter{footnote}{-1}%
  \endgroup
}

\title{Bayesian variable selection in hierarchical difference-in-differences models}
\author{James Normington$^a$, Eric F. Lock$^a$, Thomas A. Murray$^a$, and Caroline Carlin$^b$}
\date{\today}

\begin{document}
\maketitle
\blfootnote{$^a$Division of Biostatistics, School of Public Health, University of Minnesota}
\blfootnote{$^b$Department of Family Medicine and Community Health, University of Minnesota}
        
\begin{abstract}
A popular method for estimating a causal treatment effect with observational data is the difference-in-differences (DiD) model. In this work, we consider an extension of the classical DiD setting to the hierarchical context in which data cannot be matched at the most granular level (e.g., individual-level differences are unobservable).  We propose a Bayesian \textit{hierarchical difference-in-differences} (HDiD) model which estimates the treatment effect by regressing the treatment on a latent variable representing the mean change in group-level outcome. We present theoretical and empirical results showing that an HDiD model that fails to adjust for a particular class of confounding variables, or confounding with the baseline (pre-treatment) outcomes, biases the treatment effect estimate. We propose and implement various approaches to perform variable selection using a structured Bayesian spike-and-slab model in the HDiD context. Our proposed methods leverage the temporal structure within the DiD context to select those covariates that lead to unbiased and efficient estimation of the causal treatment effect. We evaluate the methods' properties through theoretical results and simulation, and we use them to assess the impact of primary care redesign of clinics in Minnesota on the management of diabetes outcomes from 2008 to 2017.  \\
\blfootnote{
    Research reported in this publication was supported by the National Institute of Diabetes and Digestive and Kidney Diseases of the National Institutes of Health under Award Number R18DK110732. The content is solely the responsibility of the authors and does not necessarily represent the official views of the National Institutes of Health.}
\end{abstract}

\section{Introduction}\label{sec:intro}
\subsection{Difference-in-differences models}\label{sec:DiD}
A common task in biostatistics is to estimate a treatment effect with observational data. A major issue in making causal inferences from observational studies is the virtually unavoidable presence of confounding variables. A popular observational method that avoids the effect of static confounders, or confounding variables whose values and relation to the outcome do not change over time, is the difference-in-differences (DiD) model. DiD estimation first defines a treatment and an outcome. Under the classical setting the treatment is binary, defining two groups (e.g., ``treatment'' and ``control'').  The standard DiD model tests for a difference between the average change in outcome over time in the treatment group and the average change in outcome over time in the control group. Specifically, it takes the difference in mean outcome between the groups \textit{before} the treatment (Difference 1), takes the difference in mean outcome between the groups \textit{after} the treatment (Difference 2), and then takes the difference between those two differences (Difference 2 - Difference 1). The DiD model is used often in econometrics, social science, and marketing. As a canonical example of applying the DiD model, Card \& Krueger compared the change in employment in New Jersey vs. Pennsylvania after New Jersey adopted an increase in the minimum wage; in this context New Jersey could be considered the treatment group and Pennsylvania the control group \cite{card1994minimum}.

Formally, let $Y_{i}^{(t)}$ denote the outcome variable for subject $i = 1, ..., n$ at timepoint $t \in \{0, 1\}$, where  $t = 0$ indicates the measurement was taken before the treatment and $t = 1$ indicates the measurement was taken after the treatment. Let $T_i^{(t)}$ denote the treatment status for individual $i$ at timepoint $t$. The observed treatment level for individual $i$ is then $T_i \equiv T_i^{(1)} - T_i^{(0)}$. In the DiD framework, with a binary treatment, the ``treatment'' group's treatment status is $T_i = T_i^{(1)} - T_i^{(0)} = 1 - 0 = 1$ and the ``control'' group's treatment status is $T_i = T_i^{(1)} - T_i^{(0)} = 0 - 0 = 0$. In the continuous case, subject $i$'s treatment status is simply $T_i = T_i^{(1)} - T_i^{(0)}$.
A common DiD model is
\begin{equation}
    Y_{i}^{(t)} = \beta_{0} + \phi \mathbbm{1}(t = 1) + \Delta T_i \mathbbm{1}(t = 1) + \epsilon_{i}^{(t)}
    \label{eq:standard_DiD}
\end{equation}
 where $\beta_{0}$ is the pre-treatment mean of the control group, $\phi$ is the common time trend assumed for each subject, $\Delta$ is the treatment effect of interest,  $\mathbbm{1}$ is the indicator function, and $\epsilon_{i}^{(t)}$ is the subject's normally distributed residual at timepoint $t$. In this formulation, $\Delta$ is the expected difference in post-treatment outcome for an individual, if they receive the treatment rather than the control. To define $\Delta$ formally, we use the potential outcomes notation introduced in the Rubin causal model \cite{rubin:1974}. Letting $Y_{i}(T_i = c)$ denote the potential outcome of subject $i$ had they received treatment $c$, $\Delta$ is defined as the average treatment effect in the post-treatment timepoint: $\Delta = \mathbb{E}\bigg[Y_{i}^{(1)}(T_{i}=1) - Y_{i}^{(1)}(T_{i}=0)  \bigg]$. 
 
    The typical DiD estimator is the observed difference between the average observed trend in the treated group and the average observed trend in the control group: $\hat{\Delta} = \bigg( \bar{Y}^{(1)}_{T_j = 1} - \bar{Y}^{(0)}_{T_j = 1} \bigg) - \bigg( \bar{Y}^{(1)}_{T_j = 0} - \bar{Y}^{(0)}_{T_j = 0} \bigg)$. If certain identifying assumptions are satisfied, then $\hat{\Delta}$ is an unbiased and consistent estimator of $\Delta$ \cite{abraham:2018}.  Using $\hat{\Delta}$ to estimate $\Delta$ differences out the effect of static confounders, eliminating the need to adjust the model for them. However, an important identifying assumption is the absence of \textit{dynamic confounders}, which are confounding variables whose values or relation to the outcome change over time. 
 
More generally, the treatments $T_i$ may be non-binary or continuous. In this setting, only one of the potential outcomes $\bigg\{Y_{i}^{(1)}(T_i=c) \ \big\vert \ c \in \mathcal{T} \bigg\}$, where $\mathcal{T}$ denotes the support of $\bT$, is observed.  Here, $\Delta$ gives the expected difference in post-treatment outcome for an individual if they receive an additional unit of exposure:
 \begin{equation}
 \Delta = \mathbb{E}\bigg[Y_{i}^{(1)}(T_{i}=c+1) - Y_{i}^{(1)}(T_{i}=c) \bigg].
    \label{eq:causal_effect}
\end{equation}

\subsection{Covariate adjustment and selection} 
\label{sec:var_adjust}
Adjusting an estimator of a causal treatment effect to eliminate confounding is a well-explored topic in the single timepoint context. Propensity score matching and inverse probability of treatment weighting are popular methods which use conditional exchangeability (i.e., conditional on some other covariate(s), treatment groups are comparable) to eliminate confounding \cite{rosenbaum:1983, horvitz:1952}. However, these methods require adjustment for \textit{all} confounding variables to estimate the treatment effect consistently. 

More recently, approaches which jointly estimate outcome and exposure/treatment models, and then leverage information across each model to estimate the treatment effect have been proposed. Wang et al. and Cefalu propose methods which identify potential confounders by imposing prior dependence on a covariate's inclusion in the propensity score and outcome models\cite{wang:2012, cefalu:2016}. Zigler and Dominici propose Bayesian methods that perform variable selection, then
estimate the treatment effect as a weighted average of estimates yielded from propensity score models with different covariates included\cite{zigler:2014}. Koch et al. propose a method which simultaneously estimates the treatment effect while performing an adaptive group lasso based variable selection algorithm \cite{koch:2017}.  

In the point-exposure context, confounding occurs when a covariate is associated with both treatment and outcome. In the DiD context, confounding occurs when (i) the covariate is associated with treatment and (iia) the association between the covariate and outcome varies over time or (iib) the covariate evolves over time differently in the treatment and control groups \cite{wing:2018}.
Thus, identifying the causal effect \eqref{eq:causal_effect} may require extending model \eqref{eq:standard_DiD} as follows:
\begin{equation}
    Y_{i}^{(t)} = \beta_{0} + \phi \mathbbm{1}(t = 1) + \Delta T_i \mathbbm{1}(t = 1) + \sum_{k=1}^K \beta_k^{(t)} X_{ik}^{(t)}+ \epsilon_{i}^{(t)}.
    \label{eq:covariate_DiD}
\end{equation}
Static covariates ($X_{ik}^{(0)} = X_{ik}^{(1)}$) may still be confounders of type (iia) if $\beta_k^{(0)} \neq \beta_k^{(1)}$;  confounders of type (iib) require  $X_{ik}^{(0)} \neq X_{ik}^{(1)}$.

Many applications of the covariate-adjusted DiD model directly select the covariates without any data-driven variable selection techniques \cite{abadie:2005, ionesco:2014, mckinnon:2015}. Stuart et al. uses propensity scores to increase comparability of the four DiD groups (treatment-pre, control-pre, treatment-post, control-post)\cite{stuart:2014}. Sofer et al. interpret the DiD model as a negative control outcome to identify and correct for biased DiD-estimated treatment effects\cite{sofer:2016}. 

An often used Bayesian variable selection approach is \textit{spike-and-slab} variable selection, which proceeds by specifying a two-component mixture prior on each regression coefficient. In George \mbox{$\&$} McCulloch's implementation, the spike and slab are mean 0 Gaussian distributions with low and high variances, respectively:
\begin{align}
\begin{split}
&    \beta_k \vert w_k, z_k^2, c^2 \sim (1-w_k) N(0, z_k^2) + w_k N(0, c^2 z_k^2),  \\
&    w_k \vert p \sim \text{Bernoulli}(p), \ k = 1, ..., K,
\end{split}
\label{eq:spike_and_slab_GE}
\end{align}
where $w_k = 1$ indicates $\beta_k$ was selected for the slab and $w_k = 0$ indicates $\beta_k$ was selected for the spike \cite{george:1993}. To make this formulation work, $z_k^2$ is chosen to be small and $c^2$ is chosen to be large. If the data suggest \mbox{$\beta_k$} is nonzero, it will have more posterior mass under the slab prior and thus its corresponding covariate will be included in the model. If the data suggest \mbox{$\beta_k$} is close to zero, it will have more posterior mass under the spike prior and thus its corresponding covariate will be excluded from the model. In this paper, we base our proposed variable selection methods on modified versions of this spike-and-slab prior.

\subsection{Hierarchical difference-in-differences models}\label{sec:hierarchical_DiD}
For model~\eqref{eq:covariate_DiD}, it is often sufficient to consider the change in outcome for each individual:
\begin{equation}
    Y_{i}^{\text{diff}} = \phi + \Delta T_i  + \sum_{k=1}^K \beta_k X_{ik}+ \epsilon_{i}
    \label{eq:covariate_DiD_change}
\end{equation}
where $Y_i^{\text{diff}} = Y_i^{(1)}-Y_i^{(0)}$ and $\epsilon_{i}= \epsilon_i^{(1)}-\epsilon_i^{(0)}$.  For static covariates (i.e., $X_{ik}=X_{ik}^{(1)}=X_{ik}^{(0)}$ so that $\beta_k = \beta_k^{(1)}- \beta_k^{(0)}$) and dynamic covariates with constant association with the outcome
(i.e., $X_{ik} = X_{ik}^{(1)}-X_{ik}^{(0)}$ so that $\beta_k = \beta_k^{(1)} = \beta_k^{(0)}$), this is a classical linear model, in which estimation and inference proceeds with well-understood theoretical results.   

We instead consider the hierarchical context in which change is not observed at the most granular level (e.g., not observed for individuals).  Our motivating example is an application to assess the impact of primary care redesign policy on diabetes outcomes at clinics in Minnesota.  Individual outcomes are not matched longitudinally, and thus the mean change in outcome, treatment(s), and covariates are measured at the clinic level.   

To circumvent these restrictions, one could alter the model in Equation~\eqref{eq:standard_DiD} to allow for multiple (potentially continuous) treatment levels (e.g., dose-response relationships) as well as adjust for group-level dynamic confounders in a hierarchical fashion:
\begin{align}
\begin{split}
& Y_{ji}^{(0)} \sim N(\mu_{j}, \widetilde{\sigma}_j^2) \\
& Y_{ji}^{(1)} \sim N(\mu_{j} + \mu_j^{\text{diff}}, \sigma_j^2) \\
& \mu_j \sim N(T_j\widetilde{\Delta}  + \bX_j^T \bBetatilde, \widetilde{\tau}^2) \\
& \mu_j^{\text{diff}} \sim N(T_j\Delta + \bX_j^T \bBeta, \tau^2),
\end{split}
\label{eq:HDiD}
\end{align}
where \mbox{$j$} is extended to \mbox{$J$} groups 
(e.g., clinics), \mbox{$j = 1, ..., J$}, $\mu_j$ is the pre-treatment mean outcome of group $j$, $\mu_j^{\text{diff}}$ is the mean change in outcome of group $j$ from $t=0$ to $t=1$, and $\bX$ is a design matrix comprised of group-level covariates with rows $\bX_j^T$. $\widetilde{\sigma}^2_j$ and $\sigma^2_j$ are group-$j$ specific outcome variances pre- and post-treatment, respectively, $T_j$ is the treatment exposure of group $j$, and $\widetilde{\tau}^2$ and $\tau^2$ are the variances of the pre-treatment mean outcome and mean change, respectively. We collect the group-level treatment exposures $T_j$ in a vector $\bT$. To measure multiple treatment effects, one could alter \mbox{Equation~\eqref{eq:HDiD}} extending the vector $\bT$ to a matrix of group-level exposures and the scalar $\Delta$ to a vector of treatment effects. We focus on estimating one treatment effect in this paper. We clarify that adjustment for $\bT$ in the third line of~\eqref{eq:HDiD} does not reflect our belief that $\bT$ drives pre-treatment values of $\bmu$, rather we include it here to control for confounding factors related to both $\bT$ and $\bmu$; this will be discussed more in depth in Section~\ref{sec:individual_bias}.

For the remainder of the paper, we refer to the third and fourth lines of Equation~\eqref{eq:HDiD} as the ``baseline'' and ``change'' models, respectively, and Equation~\mbox{\eqref{eq:HDiD}} as a \textit{hierarchical difference-in-differences model} (HDiD).
An analogous model was first introduced by Normington et al., with a discussion of required assumptions for this model to identify the causal effect~\eqref{eq:causal_effect} \cite{normington:2019}.
In this hierarchical context it is often *not* sufficient to consider change only; it is necessary to model both baseline and change ($\mu_j$ and $\mu_j^{\text{diff}}$) together  as discussed in Section~\ref{sec:causal}. 

After specifying the likelihood in Equation~\eqref{eq:HDiD}, we choose to proceed with Bayesian inference throughout this paper as the structure of Equation~\eqref{eq:HDiD} naturally lends itself to a Bayesian hierarchical model. After assigning priors, we can simply compute the posterior distribution of each parameter with standard Markov chain Monte Carlo (MCMC) techniques. Bayesian inference also facilitates incorporating prior information, if available, about model parameters, most importantly $\Delta$ in this context.

\subsection{Our contribution}
\label{contribution}

In this paper, we 
motivate and implement four variable selection methods in the HDiD framework, explore the operating characteristics therein, and apply the methods to a motivating data set. Section~\ref{sec:causal} discusses the role of confounding within the context of the model in Equation~\eqref{eq:HDiD}, suggesting which candidate variables lead to 
unbiased and efficient estimation of $\Delta$ through simulation. Section~\ref{sec:VS_in_DiD} proposes four algorithms to perform variable selection in the HDiD framework. Section~\ref{sec:simulations} conducts a simulation study to assess how each algorithm performs under a set of underlying truths. Section~\ref{sec:application} applies these variable selection techniques to study of the effect of primary care redesign policy on diabetes outcomes at participating clinics in Minnesota. Section~\ref{sec:conclusion} concludes the work and offers future directions for study in this topic.  

\section{Covariate adjustment in hierarchical difference-in-differences models}\label{sec:causal}
\subsection{Causal relationships of interest}\label{sec:causal_relationships}
Before motivating the variable selection techniques,
we first consider a simple example of the assumed data generation in Equation~\eqref{eq:HDiD}. The observed subject-level pre-treatment outcome $Y_{ji}^{(0)}$ is assumed to arise from a group $j$-specific pre-treatment mean $\mu_j$ and the observed subject-level post-treatment outcome $Y_{ji}^{(1)}$ is assumed to arise from a group $j$-specific pre-treatment mean $\mu_j$ modified by a group $j$-specific mean change $\mu_j^{\text{diff}}$. It is then assumed that $\mu_j$ and $\mu_j^{\text{diff}}$ arise from a mean structure modified by covariates through the design matrix $\bX$ and treatment vector $\bT$; this includes the treatment $\bT$ possibly affecting $\bmudiff$, as well as variables which could affect any combination of $\bT$, $\bmudiff$, and $\bmu$. 

\begin{figure}[!htbp]
    \centering
    \includegraphics[width=\textwidth]{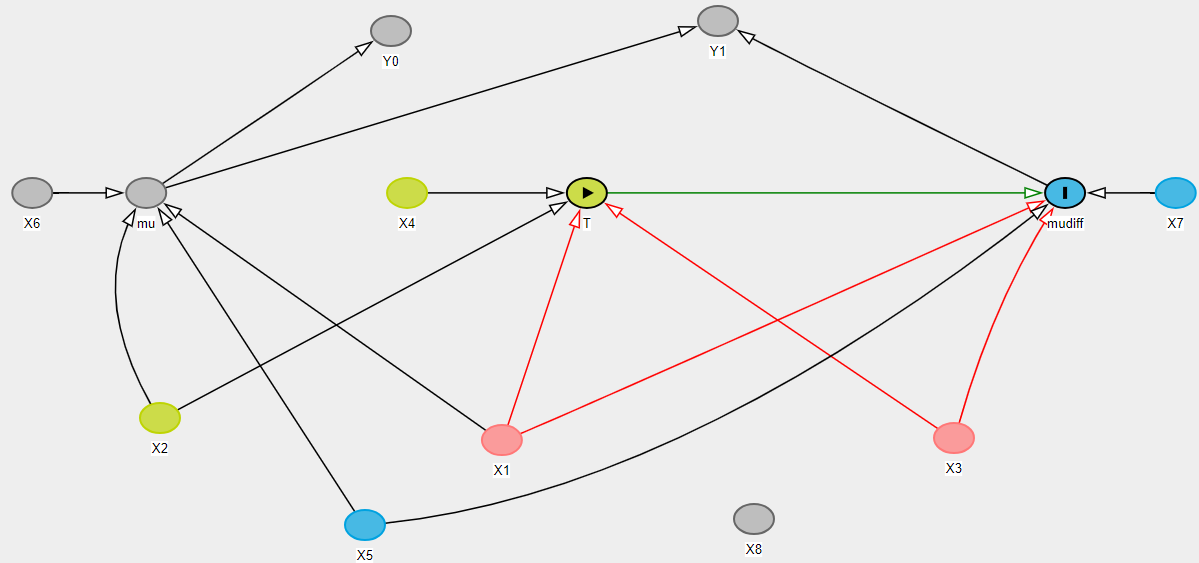}
    \caption{Assumed causal relationships and data generation throughout the paper}
    \label{fig:DAG}
\end{figure}

For our theoretical and simulation evaluations, we consider a  hypothetical set of covariates that cover each of the $2^3 = 8$ combinations of does/does not affect $\bT$, does/does not affect $\bmu$, and does/does not affect $\bmudiff$; Table~\ref{table:cases} lists these cases. Additionally, Figure~\ref{fig:DAG} displays the assumed data generation process as a directed acyclic graph \cite{pearl:1995} and was generated using the ``DAGitty'' R package \cite{textor:2016}. Our goal is to estimate the causal effect of \mbox{$\bT$} on \mbox{$\bmudiff$}, represented by the green line in Figure~\mbox{\ref{fig:DAG}}.

\subsection{Data generation}\label{sec:data_generation}
Throughout this work, we use simulation to assess the operating characteristics of our approaches. Below, we describe the data generation used in each simulation.
\begin{enumerate}
    \item Generate group-level covariate matrix $\bX$, where each column $\bX_k$ is generated as $\bX_k \overset{iid} \sim \boldsymbol{N_J}(\boldsymbol{0}, \boldsymbol{I_J})$, where $\boldsymbol{I_J}$ is the $J$-dimensional identity matrix. 
    \item Generate group-level $\boldsymbol{T} \sim \boldsymbol{N_J}(\bX \boldsymbol{\alpha}, \boldsymbol{I_J})$.
    \item Generate group-level $\bmu \sim \boldsymbol{N_J}(\bX \bBetatilde, \boldsymbol{I_J})$ and $\bmudiff \sim \boldsymbol{N_J}(\bT \Delta + \bX \bBeta, \boldsymbol{I_J})$.
    \item Generate subject-level $Y_{ji}^{(0)} \sim N(\mu_j, 1)$ and $Y_{ji}^{(1)} \sim N(\mu_j + \mu_j^{\text{diff}}, 1)$, for $j=1, ..., J$ and $i = 1, ..., n_j^{(t)}$.
\end{enumerate}
So, $\boldsymbol{\alpha}$ controls which variables are predictive of treatment, $\bBetatilde$ controls which variables are predictive of baseline mean, and $\bBeta$ controls which variables are predictive of change in means.
For each simulation, we use either $J=50$ or $J=100$ groups, each having $n_j^{(0)} = n_j^{(1)} = 10$ subjects, $j=1, ..., J$. 
For example, a generative model that includes all $8$ hypothetical covariates in Table~\ref{table:cases} and Figure~\ref{fig:DAG} is given by setting $\boldsymbol{\alpha} = [1, 1, 1, 1, 0, 0, 0, 0]$, $\bBetatilde = [1, 1, 0, 0, 1, 1, 0, 0]$, $\bBeta = [1, 0, 1, 0, 1, 0, 1, 0]$, $\widetilde{\Delta} = 0$, and $\Delta = 1$.

%

Since $\Delta$ measures the covariate-adjusted relationship between $\bT$ and $\bmudiff$, it may not be intuitive why Equation~\eqref{eq:HDiD} includes a covariate-adjusted baseline model. Even if the change model is appropriately specified, estimates of $\Delta$ may still be biased if predictors correlate with $\bmu$, $\bmudiff$, and $\bT$ but are not included in the baseline model \cite{normington:2019}. In the following sections, we measure the bias and efficiency of \mbox{$\Delta$} incurred by omitting important predictors of treatment, baseline, and change. In Section~\mbox{\ref{sec:omitted}} we apply a familiar theoretical result. In Section~\mbox{\ref{sec:individual_bias}}, we conduct a simulation to investigate the role adjusting (or not adjusting) the baseline and change models for individual covariates plays in estimating \mbox{$\Delta$}.

\subsection{Omitted variable bias}\label{sec:omitted}
After marginalizing \mbox{$\bY^{(0)}$} and \mbox{$\bY^{(1)}$} over \mbox{$\bmu$} and \mbox{$\bmudiff$} and then concatenating them into a vector \mbox{$\bY$}, the model in \mbox{\eqref{eq:HDiD}} can be re-expressed as
\mbox{$\bY \sim \boldsymbol{N(\boldsymbol{AB_1}\bTheta_1, \bSigma)}$}, where  $\boldsymbol{A} \equiv 
\begin{bmatrix}
\boldsymbol{A_0} & \boldsymbol{0} \\
\boldsymbol{0} & \boldsymbol{A_1}
\end{bmatrix}$  assigns pre- and post-treatment group-level predictors to the subject level,
$\boldsymbol{B_1} \equiv
\begin{bmatrix}
\bX_{\boldsymbol{\widetilde{w}} = 1} & \widetilde{\bT} & \boldsymbol{0} & \boldsymbol{0} \\
\bX_{\boldsymbol{\widetilde{w}} = 1} & \widetilde{\bT} & \boldsymbol{X}_{\boldsymbol{w} = 1} & \boldsymbol{T} 
\end{bmatrix}$ such that $\bX_{\boldsymbol{\widetilde{w}} = 1}$ and $\boldsymbol{X}_{\boldsymbol{w} = 1}$ are those covariates that are included in the baseline and change models (respectively) and $\widetilde{\bT}$ is $\bT$ when $\bT$ is included in the baseline model and $\boldsymbol{0}$ when it is not, and $\boldsymbol{\Theta_1} \equiv [\bBetatilde_{\boldsymbol{\widetilde{w}} = 1} \ \widetilde{\Delta} \ \bBeta_{\boldsymbol{w} = 1} \ \Delta]^T$. 
Whereas $\{\boldsymbol{B_1},\bTheta_1\}$ define the model used for estimation, assume the true generative model is \mbox{$\bY \sim \boldsymbol{N(\boldsymbol{AB}\bTheta, \bSigma)}$} where $\boldsymbol{B} \equiv [\boldsymbol{B_1} \ \boldsymbol{B_0}]$ with 
$\boldsymbol{B_0}\equiv
\begin{bmatrix}
\boldsymbol{\widetilde{U}} & \boldsymbol{0} \\
\boldsymbol{\widetilde{U}} & \boldsymbol{U} \\
\end{bmatrix}$ where $\boldsymbol{\widetilde{U}}$ and $\boldsymbol{U}$ are those covariates (with $\boldsymbol{\widetilde{U}}$  including $\bT$ when not adjusting $\bmu$ for $\bT$) excluded from the baseline and change models (respectively), and $\bTheta \equiv [\bTheta_1 \ \bTheta_0]^T$ where $\bTheta_0$ is a vector of coefficients corresponding to those covariates excluded from the model. We can then apply a modified version of the familiar omitted-variable bias result \cite{greene:2003}:

\begin{theorem*}
Suppose $\bY \sim \boldsymbol{N}(\boldsymbol{AB}\bTheta, \bSigma)$, where $\boldsymbol{B}$ is a matrix and $\bTheta$ is a vector, each which can be partitioned into $[\boldsymbol{B_1} \ \boldsymbol{B_0}]$ and $[\bTheta_1 \ \bTheta_0]^T$ respectively, and $\boldsymbol{A}$ is a matrix of conforming dimension. Suppose $\bY$ is modeled as $\bY \sim \boldsymbol{N}(\boldsymbol{AB_1}\bTheta_1, \bSigma)$. Finally, let $\bTheta_1$ have a flat prior. Then, the bias of $\widehat{\bTheta}_1$ as estimated by the model is \\
\begin{equation*}
\mathbb{E}_{\bY \vert \bTheta}[\widehat{\bTheta}_1 - \bTheta_1] = 
(\boldsymbol{B_1}^T \boldsymbol{A}^T \boldsymbol{\Sigma}^{-1} \boldsymbol{AB_1})^{-1}\boldsymbol{B_1}^T \boldsymbol{A}^T \boldsymbol{\Sigma}^{-1} \boldsymbol{AB_0} \bTheta_0.
\end{equation*}
\label{theorem:omitted_bias}
\end{theorem*}
A proof is available in Appendix~\mbox{\ref{sec:appendixA}}.
While the bias of $\Delta$ can be obtained from Theorem 1, it is difficult to isolate and express it in a simple closed form. Appendix~\mbox{\ref{sec:appendixA}} also describes two simulations which compute empirical means of the bias result when the baseline and change models are adjusted for the same covariates (\mbox{$\bX_{\boldsymbol{\widetilde{w}} = 1} = \bX_{\boldsymbol{w} = 1}$}). The first simulation shows that when the baseline model is not adjusted for $\bT$, the estimation of $\Delta$ is only unbiased when the covariate sets are $\{\bX_1, \bX_2, \bX_3 \}$ from Table~\ref{table:cases}. The second simulation does adjust the baseline model for $\bT$, and shows that the estimation of $\Delta$ is unbiased when the covariates sets are $\{\bX_1, \bX_3 \}$.
As $\bX_2$ is a static confounder (see Table~\ref{table:cases}), an important consequence is that the baseline model must be adjusted for $T$ to avoid bias due to unobserved static confounders.  

\subsection{Variable-by-variable simulations}
\label{sec:individual_bias}
To isolate the impact of each combination of does/does not affect \mbox{$\bT$}, does/does not affect \mbox{$\bmu$}, and does/does not affect \mbox{$\bmudiff$} has on the estimation of \mbox{$\Delta$}, 
for various modeling choices, we conduct a simulation.
We define eight data generation scenarios by the process described in Section~\ref{sec:data_generation}; each scenario uses only one of the covariates described in Table~\ref{table:cases} as the single true generating covariate and (depending on the modeling choice) the single model covariate. Then, we estimate \mbox{$\Delta$} according to each of the following eight model choices:
\begin{itemize}
\item Choice 1: Do not adjust $\boldsymbol{\mu}$ or $\boldsymbol{\mu}^{\textbf{diff}}$ for $\boldsymbol{X}_k$.
\item Choice 2: Adjust $\boldsymbol{\mu}$, but do not adjust $\boldsymbol{\mu}^{\textbf{diff}}$, for $\boldsymbol{X}_k$.
\item Choice 3: Adjust $\boldsymbol{\mu}^{\textbf{diff}}$, but do not adjust $\boldsymbol{\mu}$, for $\boldsymbol{X}_k$.
\item Choice 4: Adjust $\boldsymbol{\mu}$ and $\boldsymbol{\mu}^{\textbf{diff}}$ for $\boldsymbol{X}_k$.
\item Choice 5: Adjust $\boldsymbol{\mu}$ for $\boldsymbol{T}$, do not adjust $\boldsymbol{\mu}^{\textbf{diff}}$ for $\boldsymbol{X}_k$.
\item Choice 6: Adjust $\boldsymbol{\mu}$ for $\boldsymbol{T}$ and $\boldsymbol{X}_k$, but do not adjust $\boldsymbol{\mu}^{\textbf{diff}}$, for $\boldsymbol{X}_k$.
\item Choice 7: Adjust $\boldsymbol{\mu}$ for $\boldsymbol{T}$, adjust $\boldsymbol{\mu}^{\textbf{diff}}$ for $\boldsymbol{X}_k$.
\item Choice 8: Adjust $\boldsymbol{\mu}$ for $\boldsymbol{T}$ and $\boldsymbol{X}_k$, adjust $\boldsymbol{\mu}^{\textbf{diff}}$ for $\boldsymbol{X}_k$.
\end{itemize}

Table~\ref{table:individual_sim} displays the bias, mean squared error (MSE), and coverage rates of \mbox{$\widehat{\Delta}$} for each data-generating covariate \mbox{$\bX_k$} under each scenario, with 5000 replications. Bolded quantities indicate optimal values; that is, estimated biases within a margin of error from 0, lowest MSEs within a margin of error of each other, and coverage rates no more than a margin of error less than 0.95. From these results, we can make the following conclusions:

\begin{itemize}
    \item Estimation of \mbox{$\Delta$} is biased when we fail to adjust $\bmudiff$ for covariates predictive of \mbox{$\bT$} and \mbox{$\bmudiff$} (here, Choices 1, 2, 5, and 6 in the \mbox{$\bX_1$} and \mbox{$\bX_3$} panels).
    \item Choices 3, 4, and 8 always lead to unbiased estimation of \mbox{$\Delta$} 
    (however, these all require that we observe the covariate $X_k$).  
    \item Estimates of \mbox{$\Delta$} are biased when we do not adjust \mbox{$\bmu$} for covariates predictive of \mbox{$\bT$} and \mbox{$\bmu$} (here, Choice 1 in the \mbox{$\bX_2$} panel). This bias disappears when we include \mbox{$\bT$} as a predictor for \mbox{$\bmu$} (Choices 5 and 6). However, adjusting \mbox{$\bmudiff$} for covariates predictive of \mbox{$\bmu$} and \mbox{$\bT$} (here, \mbox{$\bX_1$} and \mbox{$\bX_3$}) without adjusting \mbox{$\bmu$} for those \mbox{$\bX_k$} (Choice 7) leads to biased estimates of \mbox{$\Delta$}.
    \item The most efficient model (by lowest-MSE criterion) varies across covariate cases; in some cases only \mbox{$\bmu$} should be adjusted, in some only \mbox{$\bmudiff$} should be adjusted, in some they should both be adjusted, and in others adjusting either needlessly complicates the model.
    \item  Choices that yield unbiased estimates of $\mbox{$\Delta$}$ have nominal coverage rates, while those with bias do not.  
\end{itemize}

In Table~\mbox{\ref{table:individual_sim}}, when there are covariates predictive of both $\bmu$ and $\bT$ (here, $\bX_1$ and $\bX_2$), the bias of \mbox{$\widehat{\Delta}$} is decreased when \mbox{$\bT$} is included as a predictor for \mbox{$\bmu$} (Choice 1, Choice 5). However, when such a covariate is excluded from \mbox{$\bmu$}, \mbox{$\bT$} is retained as a predictor for \mbox{$\mu$}, and the covariate is included for \mbox{$\bmudiff$} (Choice 7), there is noticeable bias. To avoid this bias, one could force the baseline and change models to include the same predictors.  Another approach to avoid this bias would instead be to include those \mbox{$\bX_k$} in the change model that are predictive of \mbox{$\bT$} and \mbox{$\bmudiff$}, and then only include those \mbox{$\bX_k$} in the baseline model that are predictive of \mbox{$\bT$}, \mbox{$\bmudiff$}, and \mbox{$\bmu$}.

Variable selection should also be concerned with efficient estimation. When there are covariates predictive of \mbox{$\bmu$} but not \mbox{$\bmudiff$} (here, $\bX_2$ and $\bX_6$), bias and MSE are minimized when $\bX_k$ is included in the model for $\bmu$ but not $\bmudiff$ (Choices 2 and 6). When there are covariates predictive of \mbox{$\bmudiff$} but not \mbox{$\bmu$} (here, \mbox{$\bX_3$} and \mbox{$\bX_7$}), MSE is minimized when \mbox{$\bX_k$} is included in the model for \mbox{$\bmudiff$} regardless of the baseline model adjustment (Choices 3, 4, 7, and 8). One approach to variable selection, then, is to perform variable selection on the baseline and change models separately. To arrive at a more parsimonious (i.e., fewer covariates used) model, another alternative approach is to include any \mbox{$\bX_k$} predictive of \mbox{$\bmudiff$} in the change model, and only include any \mbox{$\bX_k$} predictive of \mbox{$\bmu$} \textit{and} \mbox{$\bmudiff$} in the baseline model.
 
\section{Variable selection approaches}\label{sec:VS_in_DiD}
Motivated by the results in Section~\ref{sec:causal}, we propose four variable selection approaches using the Bayesian spike-and-slab prior in \mbox{\eqref{eq:spike_and_slab_GE}}. The slab component of this prior is not limited to be Gaussian. In fact, Ghosh et al. recommend a central t-distribution, which has good support for values of $\beta_k$ moderately far from 0, but unlike a Gaussian distribution it has heavier tails to support values of $\beta_k$ very far from 0\cite{ghosh:2018}. To implement this, we modify Equation~\eqref{eq:spike_and_slab_GE} to 
\begin{align}
\begin{split}
&    \beta_k \vert w_k, z_k^2, \lambda_k \sim (1-w_k) N(0, z_k^2) + w_k t_{\nu}(0, \lambda_k),  \\
& w_k \vert p \sim \text{Bernoulli}(p), \ k = 1, ..., K, 
\end{split}
\label{eq:spike_and_slab_t}
\end{align}
where $t_{\nu}(0, \lambda_k)$ denotes the central Student t-distribution with $\nu$ degrees of freedom and scale $\lambda_k$.

To facilitate computation, we can take advantage of the t-distribution's equivalence with a scale mixture of Gaussian distributions, re-expressing Equation~\eqref{eq:spike_and_slab_t} as 
\begin{align}
\begin{split}
& \beta_k \vert w_k, z_k^2, \gamma_k \sim (1 - w_k) N(0, z_k^2) + w_k N(0, 1 / \gamma_k), \\
& \gamma_k \vert \lambda_k \sim \text{Gamma}(\text{shape = }\nu/2, \text{rate = } (\nu/2)\lambda_k^2) \\
& w_k \vert p \sim \text{Bernoulli}(p), \ k = 1, ..., K, 
\end{split}
\label{eq:spike_and_slab_t_computation}
\end{align}
with a diffuse mean-zero Gaussian prior for the intercepts $\widetilde{\beta}_0$ and $\beta_0$: $\pi(\widetilde{\beta}_0) = N(0, \widetilde{\omega}^2)$ and $\pi(\beta_0) = N(0, \omega^2)$, where $\widetilde{\omega}^2$ and $\omega^2$ are set to large constants.
 We now propose four applications of the spike-and-slab prior in Equation~\eqref{eq:spike_and_slab_t} to perform variable selection in the DiD context.
 Appendix~\ref{sec:appendixB} gives the precise Gibbs sampling algorithm for each method.

 \subsection{Separate method}\label{sec:separate}
 The first approach we consider, called the Separate method, performs variable selection separately for the baseline and change models. That is, the posterior of $\widetilde{\beta}_k$ informs whether or not to include $\bX_k$ in the model for $\bmu$, and the posterior of $\beta_k$ independently informs whether or not to include $\bX_k$ in the model for $\bmudiff$. This approach makes sense when there are some candidate variables related to either the baseline or change models, but perhaps not both. Using this method, the priors for $\widetilde{\beta}_k$ and $\beta_k$ 
 are
\begin{align}
\begin{split}
\widetilde{\beta}_k \vert \widetilde{w}_k, \widetilde{\lambda}_k \sim (1 - \widetilde{w}_k)N(0, z_k^2) + \widetilde{w}_k t_{\nu}(0, \widetilde{\lambda}_k), k = 1, ..., K    \\
\beta_k \vert w_k, \lambda_k \sim (1 - w_k) N(0, z_k^2) + w_k t_{\nu}(0, \lambda_k), k = 1, ..., K  
\end{split}
\label{eq:SnS_separate}
\end{align}
where $\widetilde{w}_k$ is 1 if $\bX_k$ is included in the baseline model and 0 otherwise, and $w_k$ is 1 if $\bX_k$ is included in the change model and 0 otherwise.  
We can re-express~\eqref{eq:SnS_separate} as a multivariate Gaussian distribution using the scale mixture definition of the t-distribution in~\eqref{eq:spike_and_slab_t_computation}. Specifically, we define $a_k = 1/\sqrt{\gamma_k}$ when $w_k = 1$ and $a_k = z_k$ if $w_k = 0$. Then, the prior for $\bBeta$ arises as $\bBeta \sim \boldsymbol{N} (\boldsymbol{0}, \boldsymbol{D}^2)$, where $\boldsymbol{D} \equiv \text{diag}(\omega, a_1, ..., a_K)$. The setup for $\bBetatilde$ is analogous:  $\bBetatilde \sim \boldsymbol{N} (\boldsymbol{0}, \boldsymbol{\widetilde{D}}^2)$.

\subsection{Shared method}\label{sec:shared}
While the Separate method makes sense when variables are clearly related to either the baseline model or the change model (but not both), there may be scenarios in which some covariates are predictive of both. For example, consider a county-level initiative to encourage its residents to recycle more. If the average socioeconomic status of the county were positively related to baseline per-capita pounds recycled, one would want to include the county's socioeconomic status as a covariate in the baseline model. If residents with higher socioeconomic status were also more willing to change their recycling habits, one would also want to include socioeconomic status in the change model. In this instance, the Separate method would use the posterior masses of $\widetilde{\beta}_k$ and $\beta_k$ separately to make independent draws for $\widetilde{w}_k$ and $w_k$, where the optimal approach in this instance would be to use information from both posteriors to strengthen the probability of including a covariate that should be included in both models. 

The second approach, which we call the Shared method, is to constrain $\bX_k$ to be either excluded or included in both models based on the joint posterior distributions of $\widetilde{\beta}_k$ and $\beta_k$. We specify a joint prior for [$\widetilde{\beta}_k \ \beta_k$]$^T$ with a shared inclusion indicator $w_k$: 
\begin{align}
\begin{split}
& [\widetilde{\beta}_k \ \beta_k]^T \ \vert \ w_k, \lambda_k \sim (1 - w_k) \boldsymbol{N_2}(\boldsymbol{0}, z_k^2\boldsymbol{I}_2) + w_k \boldsymbol{t}_{\nu}(\boldsymbol{0}, \lambda_k \boldsymbol{I}_2), k = 1, ...K \\ 
\end{split}
\label{eq:SnS_shared}
\end{align}
where $\boldsymbol{t}_{\nu}$ is a bivariate t-distribution with $\nu$ degrees of freedom. 

\subsection{Sufficient method}
The Separate and Shared methods provide approaches which include $\bX_k$ if there is statistical evidence that $\bX_k$ reduces the residual variance in $\bmu$ and/or $\bmudiff$. These approaches are especially sensible when the number of groups $J$ is small, wherein inference on $\Delta$ heavily depends on which covariates are included in the models. Suppose instead that the number of groups $J$ is large, so that estimates of $\Delta$ are relatively precise. In this scenario, the only covariates that need to be included to estimate $\Delta$ without bias are those related to both $\bT$ and $\bmudiff$. For example, Table~\ref{table:individual_sim} suggests that when we adjust the baseline model for \mbox{$\bT$} (Choices 5-8), the minimally sufficient set of covariates needed to estimate $\Delta$ without bias are $\{\bX_1, \bX_3\}$. We propose a third method, which we call the Sufficient method, to identify the smallest model that allows unbiased estimation of $\Delta$. 

First, we introduce an exposure-confounder model to identify those $\bX_k$ that are associated with $\boldsymbol{T}$. If $\bT$ is binary, this may be a probit or logistic model. If instead $\bT$ is continuous, a reasonable exposure model may be
\begin{equation}
    \boldsymbol{T} = \bX \boldsymbol{\alpha} + \boldsymbol{\epsilon}_{\alpha}, \boldsymbol{\epsilon}_{\alpha} \sim \boldsymbol{N}(\boldsymbol{0}, \sigma^2_{\alpha} \boldsymbol{I})
    \label{eq:exposure_confounder_linear}
\end{equation}
with a noninformative prior on $\sigma^2_{\alpha}$: $\pi(\sigma^2_{\alpha}) \propto 1 / \sigma^2_{\alpha}$. Then, we can impose a similar spike-and-slab prior on each $\alpha_k$:
\begin{equation*}
    \alpha_k \vert w_k^e, \lambda_k^e \sim (1-w_k^e)N(0, z_k^e) + w_k^e t_{\nu}(0, \lambda_k^e) 
\end{equation*}
with $\pi(w_k^e) = \text{Bern}(p^e)$, where $w_k^e = 1$ when $\bX_k$ is selected for this exposure model and is 0 otherwise. 

To only include those covariates for $\bmudiff$ that are also predictive of $\bT$, we set $w_k=0$ if $w_k^e = 0$; that is, $\pi(w_k) = w_k^e*\text{Bern}(p)$. 
Similarly, to only include those covariates for $\bmu$
that are also predictive of $\bmu$, $\bmudiff$, and $\bT$, we only allow $\widetilde{w}_k$ to be 1 when $w_k = 1$: $\pi(\widetilde{w}_k) = w_k*\text{Bern}(\widetilde{p})$. 

Given the exposure model in~\mbox{\eqref{eq:exposure_confounder_linear}} and outcome model in~\mbox{\eqref{eq:HDiD}}, estimation of the treatment effect
usually proceeds by first estimating $\boldsymbol{\alpha}$ using the exposure model, treating estimates of $\boldsymbol{\alpha}$ as fixed and known, and then estimating $\Delta$ using the outcome model. An alternative fully multivariate approach is to combine~\mbox{\eqref{eq:exposure_confounder_linear}} and \mbox{\eqref{eq:HDiD}} into one joint likelihood and model both simultaneously in a Bayesian framework. However, Zigler et al. show that such an approach can lead to ``feedback'' between the two models; that is, quantities in the outcome model informing quantities in the exposure model\cite{zigler:2013}. In general, model feedback can lead to biased estimates of the treatment effect. To prevent this, we choose to perform variable selection sequentially across the exposure, change, and baseline models. Specifically, within each Gibbs iteration,
 we draw $w_k^e$ without conditioning on the baseline or change models, and we draw $w_k$ without conditioning on the baseline model.  

\subsection{Efficient method}
Table~\ref{table:individual_sim} suggests that the most 
efficient estimates of $\Delta$, 
in terms of MSE, occur when the covariate set includes 
those $\bX_k$ related to $\bmudiff$ and excludes those $\bX_k$ that are not. 
Combining this with the desire for model parsimony, we blend the Separate and Sufficient methods to include only those $\bX_k$ related to $\bmudiff$. To start, we use the same spike-and-slab prior as in Equation~\eqref{eq:SnS_separate} and do not fit an exposure model. Since the decision to include $\bX_k$ in the model for $\bmudiff$ depends on whether $\bX_k$ predicts $\bT$, we assign a prior for $w_k$ that is independent of the exposure model: $\pi(w_k) = \text{Bern}(p)$. When selecting covariates for the baseline, it suffices to consider only those that are also predictive of change, so we only allow $\widetilde{w}_k$ to be 1 when $w_k = 1$: $\pi(\widetilde{w}_k) = w_k*\text{Bern}(\widetilde{p})$. Again to prevent model feedback, we draw $w_k$ without conditioning on the model for baseline.  

\section{Simulations to assess variable selection algorithms}
\label{sec:simulations}
We conducted a simulation study to assess bias, MSE, and coverage rates of $\Delta$ when implementing the Separate, Shared, Sufficient, and Efficient methods using the covariate set in Section~\ref{sec:causal_relationships} and data generation scheme described in Section~\ref{sec:data_generation}, with $\boldsymbol{\alpha} = [1, 1, 1, 1, 0, 0, 0, 0]$, $\bBetatilde = [1, 1, 0, 0, 1, 1, 0, 0]$, $\bBeta = [1, 0, 1, 0, 1, 0, 1, 0]$, $\widetilde{\Delta} = 0$, and $\Delta = 1$.

For the slab, we chose a t-distribution with $\nu=5$ degrees of freedom and $\lambda_k = 5$, which has regression coefficients moderately far from 0, and has thick tails to support $\beta$ very far from 0. To allow for conditionally conjugate Gibbs sampling, we re-expressed the slabs for $\widetilde{\beta}_k$ and $\beta_k$ in \eqref{eq:SnS_separate} as $N(0, 1 / \widetilde{\gamma}_k)$ and $N(0, 1 / \gamma_k)$, with weakly informative $\text{Gamma}(\text{shape = }5/2, \text{rate = }(5/2)*5^2)$ priors for each $\widetilde{\gamma}_k$ and $\gamma_k$. Finally, we chose a spike variance of $z_k^2 = 0.01^2$, prior inclusion probabilities $\widetilde{p} = p = p_e = 1/2$, and prior variance for the intercept coefficients $\widetilde{\omega}^2 = \omega^2 = 10000$.

In the Shared approach, we again took advantage of the scale mixture of Gaussian distributions to represent the t-distribution by re-expressing Equation~\eqref{eq:SnS_shared} as
\begin{align*}
\begin{split}
& [\widetilde{\beta}_k \ \beta_k]^T \ \vert \ w_k, \gamma_k \sim (1 - w_k) \boldsymbol{N_2}(\boldsymbol{0}, 0.01^2\boldsymbol{I}_2) + w_k \boldsymbol{N_2}(\boldsymbol{0}, 1/\gamma_k \boldsymbol{I}_2), k = 1, ...K \\ 
\end{split}
\label{eq:SnS_shared_computation}
\end{align*}

with $\pi(\gamma_1) = ... = \pi(\gamma_K) = \text{Gamma}(5/2, (5/2)*5^2)$.

Tables~\ref{table:study2_J50}-\ref{table:inclusion_mudiff_J100} display, for number of groups $J=50$ and $J=100$, the bias, MSE, and coverage rates of $\Delta$, the mean number of predictors included in the change and baseline models, and the inclusion probabilities for each predictor. The results are presented for each variable selection algorithm 
with the results using the ``Full'' model (with covariates $\{\bX_1, ..., \bX_8 \}$ for $\bmu$ and $\bmudiff$) and the ``Null'' model (with no covariates for either model). Overall, the Separate and Efficient methods performed the best in terms of bias and MSE. The Sufficient method suffered higher bias and MSE with lower coverage; a brief investigation showed 
that this was due to uncertainty in the variable selection for the exposure model. The Null model performed poorly, having the highest bias and MSE with unacceptably low coverage.
While $\hat{\Delta}$ as estimated by the Full model was unbiased, it had higher MSE than the methods with variable selection in almost every scenario.  The Shared method included the most predictors for $\bmudiff$, while the Sufficient method yielded the most parsimonious model for $\bmudiff$. Tables~\ref{table:inclusion_mudiff_J50} and \ref{table:inclusion_mudiff_J100} show that the inclusion probabilities behave as desired using each method. Specifically, the Separate and Efficient methods included covariates that were predictive of \mbox{$\bmudiff$} often, the Shared method included covariates that were predictive \mbox{$\bmudiff$} or \mbox{$\bmu$} often, and the Sufficient method only included those covariates that were predictive of \mbox{$\bmudiff$} and \mbox{$\bT$} often.
The bias and MSE (with their margins of error) decreased as $J$ increased from 50 to 100, suggesting that each variable selection method leads to consistent estimation of $\Delta$. Similarly, the inclusion probabilities approached either 0 or 1 as $J$ increased, suggesting that the Separate, Sufficient, and Efficient methods are consistent in including the variables that they are designed to select. The Shared method tended to include a variable in both models if it is predictive of either baseline ($\widetilde{\beta}_k$) or change ($\beta_k$).  

\section{Application to primary care redesign and diabetes data}\label{sec:application}

Diabetes mellitus is a chronic condition that impacts the way the body breaks down glucose, its main source of energy. Diabetes can negatively impact one's endocrine, excretory, digestive, kidney, circulatory, integumentary, central nervous, and reproductive systems \cite{effects_on_body}. To identify trends in a diabetes patient's blood sugar levels, the American Diabetes Association recommends routine A1c tests and consultations with their primary care physician \cite{ADA:2018}. Primary care practices provide a structured system for individuals to manage their diabetes \cite{berwick:2008, blackwell:2012}. In 2008, the Minnesota State Legislature endorsed the Patient Centered Medical Home (PCMH) as the preferred model for primary care redesign \cite{minnesota_statutes}. 

The MN Community Measurement (MNCM) Optimal Diabetes Care dataset contains patient-level summaries of how a patient at a participating clinic managed her diabetes in a given year. It contains a patient's latest A1c, low-density lipoprotein (LDL), and systolic blood pressure (SBP) measurements for the year, as well as information on whether or not a patient is diagnosed with ischemic vascular disease (IVD), whether they have type 1 diabetes or not (Type 1), whether they have private insurance or not (Commercial), and whether or not they use tobacco (Tobacco). The American Community Survey (ACS) is an annual survey conducted by the U.S. Census Bureau which administers a questionnaire to a sample of addresses capturing many of the variables included in the long form decennial census. It contains metrics of the socioeconomic statuses of a patient's neighborhood which we used to compute composite measures of ``Wealth'' and ``Income'' and matched to the patient record at the ZIP code level \cite{swaney:2018}. 

The Physician Practice Connections-Research Survey \cite{solberg2013medical} is a survey designed to measure primary care organizational infrastructure across five of the six domains of Bodenheimer and Wagner's Chronic Care Model (CCM) \cite{bodenheimer}: health care organization, delivery system redesign, clinical information systems, decision support, and self-management support. Clinics in Minnesota were asked to report organizational structure at present (2011) and (by recall) in 2008. The PPCRS was administered a second time in 2017 to a practice population including the original health care home cohort. We identified one principal component \cite{jolliffe1986principal} driving the variance in the 2008 and 2017 PPCRS results, so we define a clinic's ``score'' as the first principal component of the 
survey matrix. To measure how a clinic matured in its primary care delivery from 2008 to 2017, we define our main exposure of interest as the clinic score difference: $\boldsymbol{c}^{\textbf{diff}} \equiv \boldsymbol{c}^{\textbf{(2017)}} - \boldsymbol{c}^{\textbf{(2008)}}$. Clustering analysis and a  histogram of $\boldsymbol{c}^{\textbf{diff}}$ implied that the higher a clinic's clinic score is, the more mature it is in its primary care transformation toward a PCMH.

In this section, we use outcome and demographic patient data from MNCM, neighborhood-level covariates from the ACS, and clinic resources and services survey data from the Physician Practice Connections-Research Survey to quantify the causal impact of primary care redesign on mean diabetes outcomes from the year 2008 (pre-treatment) to the year 2017 (post-treatment) on $J=96$ clinics. Specifically, for each outcome separately, we fit the HDiD model in~\mbox{\eqref{eq:HDiD}} and apply each of the four variable selection techniques introduced in Section~\mbox{\ref{sec:VS_in_DiD}}. Each candidate predictor is a change in proportions or means from 2008 to 2017; for example, the ``Age'' predictor for clinic $j$ is its mean patient age in 2008 subtracted from its mean patient age in 2017. In general, one should not adjust these models for post-treatment covariates that are affected by the treatment (that is, those $\bX_k$ such that $\bT \rightarrow \bX_k \rightarrow \bmudiff$). Fortunately, this is not a concern in our context. There is little contextual evidence to suggest that a clinic's speed to adopt the PCMH model affects clinic-level demographic changes. The choice to adopt the PCMH model has a small or nonexistent effect on a patient's choice of clinic, a choice probably based more on static variables such as the patient's geographic location and insurance type. For each method, we use a spike standard deviation of $z_k = 0.025$ for each candidate predictor. 

Table~\ref{table:inclusion} displays the inclusion probabilities within the change model for each candidate predictor and outcome. Change in percent of Female patients seems to be predictive of baseline A1c and LDL but not change in A1c or LDL, evidenced by the starkly different inclusion probabilities between the Shared methdods and the others. Change in percent of patients with IVD is a strong predictor of change in LDL, though must not be a strong predictor of $\boldsymbol{c}^{\textbf{diff}}$, seen in the differences in inclusion probabilities between the Sufficient and Separate, Shared, and Efficient methods. There does not appear to be much evidence that baseline clinic score $\boldsymbol{c}^{\textbf{(2008)}}$ is predictive of change in A1c, LDL, or SBP.

Tables~\ref{table:CIs_A1c}-\ref{table:CIs_SBP} display, using each variable selection method and fitting the full model, the 95$\%$ credible intervals for $\bBeta$ with the A1c, LDL, and SBP outcomes (respectively). Overall, the variable selection methods led to similar estimates 
for $\boldsymbol{c}^{\textbf{diff}}$,  a sensible result given the large number of clinics $J$ and that all methods are designed to give unbiased estimates of the treatment effect. All credible intervals for the A1c outcome contain 0. The main exposure of interest, $\boldsymbol{c}^{\textbf{diff}}$ is significantly negatively associated with the $\bmudiff$ for the LDL and SBP outcomes, suggesting that greater strides in PCMH redesign lower diabetes patients' cholesterol and blood pressure. The only other significant dynamic predictor is change in proportion of patients with IVD for the change in LDL outcome, suggesting that clinics who gained more (or retained fewer) patients with IVD from 2009 to 2017 saw the average LDL of their patients decrease. 

\section{Conclusions and future directions}\label{sec:conclusion}
In this work, we introduced a hierarchical extension of the difference-in-differences model and suggested variable selection methods therein. We showed that estimation of the treatment effect is biased if we do not adjust a baseline model either for the treatment, or for all covariates predictive of both the treatment and baseline. We then showed that in order to estimate the treatment effect without bias, we also need to adjust the change model for all covariates that jointly affect both the treatment and change in outcome. 
Covariates that are correlated with the change in means should also be included in the model for change to achieve more precise estimates of the treatment effect.

With these guidelines in place, we presented four Bayesian variable selection techniques that can be implemented in the HDiD framework. Just as in our application in Section~\ref{sec:application}, such methods are useful when subjects cannot be matched from pre- to post-treatment timepoints and when the treatment is administered at the group level. Through simulation, we found that each approach leads to reasonable estimation of the treatment effect, and the results suggest that the approaches are asymptotically unbiased. Our simulations suggested that the Sufficient method leads to the smallest covariate set able to estimate the treatment effect without bias (asymptotically), while the Efficient method estimates the treatment effect without bias (asymptotically) and with the lowest variance. We applied these methods and show that as a clinic matures in its transformation as a patient-centered medical home, the average LDL and SBP of its patients decrease. 

In this paper, we do not consider the consequences of adjusting for post-treatment covariates whose values are affected by the treatment (that is, those $\bX_k$ such that $\bT \rightarrow \bX_k \rightarrow \bmudiff$), which is extensively cautioned against in the literature \cite{rosenbaum1984consequences, montgomery:2017}. We do not recommend applying our variable selection algorithms using such covariates. Our model explicitly assumes a homogeneous treatment effect across treatment groups; this differs from the typical DiD setting in which the primary estimand of interest is the average treatment effect on the treated \cite{lechner:2010}.

Our current approaches assume a somewhat strict prior correlation structure between the inclusion indicators. Specifically, the Separate method assumes a priori that the inclusion indicators for the baseline and change model are uncorrelated, while the Shared method forces the two to be identical. The Sufficient and Efficient methods induce correlation by forcing a covariate excluded from the change model to be excluded from the baseline model. Models that allow for a more flexible dependence structure between the baseline and change models are a direction of future research.

\section*{Software}
The codes to perform the simulation in Section~\ref{sec:individual_bias} and the variable selection algorithms in Section~\ref{sec:VS_in_DiD} are available on Github: \url{https://github.com/JamesNormington/VS_in_HDiD}.

\bibliography{bibliography}{}
\bibliographystyle{SageV}

\pagebreak
\begin{table}[!htbp]
    \centering
    \begin{tabular}{c|cccccccc}
 \hline    & $\bX_1$ & $\bX_2$ & $\bX_3$ & $\bX_4$ & $\bX_5$ & $\bX_6$ & $\bX_7$ & $\bX_8$ \\
\hline \hline    $\bT$ & \checkmark & \checkmark & \checkmark & \checkmark & & & & \\
    $\bmu$ & \checkmark & \checkmark & & & \checkmark & \checkmark & &   \\
    $\bmudiff$ & \checkmark &  & \checkmark &  & \checkmark &  & \checkmark & \\
    \hline \hline
    \end{tabular}
    \caption{Role of each covariate in the causal graph}
    \label{table:cases}
\end{table}

\begin{table}[!htbp]
    \centering
    \scalebox{0.82}{
    \begin{tabular}{r|ccc|r|ccc}
\hline Choice  & Bias & MSE & Coverage  & Choice  & Bias & MSE & Coverage \\ 
    \hline $X_1$ & & & & $X_5$ & & & \\
    \hline Choice 1 & 0.528 & 0.297 & 0.038 & Choice 1 & $\boldsymbol{-0.003}$ & 0.048 & $\boldsymbol{0.949}$\\
    Choice 2 & 0.500 &  0.260 & 0.050 & Choice 2 & $\boldsymbol{0.002}$ & 0.045 & $\boldsymbol{0.951}$ \\
    Choice 3 &$\boldsymbol{0.000}$ & $\boldsymbol{0.025}$ & $\boldsymbol{0.955}$ & Choice 3 &$\boldsymbol{0.000}$ & $\boldsymbol{0.025}$ & $\boldsymbol{0.954}$ \\
    Choice 4 & $\boldsymbol{0.000}$ & $\boldsymbol{0.026}$ & $\boldsymbol{0.951}$ & Choice 4 &$\boldsymbol{0.000}$ & $\boldsymbol{0.026}$  & $\boldsymbol{0.952}$\\
    Choice 5 & 0.502 & 0.269 & 0.054 & Choice 5 & $\boldsymbol{0.002}$ & 0.046 & $\boldsymbol{0.956}$\\
    Choice 6 & 0.499 & 0.267 & 0.051 & Choice 6 & $\boldsymbol{-0.002}$ & 0.046 & $\boldsymbol{0.949}$\\
    Choice 7 & -0.041 & 0.028 & 0.943 & Choice 7 & $\boldsymbol{-0.001}$ & $\boldsymbol{0.026}$ & $\boldsymbol{0.951}$ \\
    Choice 8 & $\boldsymbol{0.001}$ & $\boldsymbol{0.025}$ & $\boldsymbol{0.948}$ & Choice 8 & $\boldsymbol{0.001}$ & $\boldsymbol{0.025}$ & $\boldsymbol{0.949}$ \\
    \hline $X_2$ & & & & $X_6$ & & & \\
    \hline Choice 1 & 0.029 & $\boldsymbol{0.013}$ & 0.935 & Choice 1 & $\boldsymbol{-0.003}$ & $\boldsymbol{0.025}$ & $\boldsymbol{0.946}$ \\
    Choice 2 & $\boldsymbol{0.002}$ & $\boldsymbol{0.013}$ & $\boldsymbol{0.952}$ & Choice 2 & $\boldsymbol{0.001}$ & $\boldsymbol{0.024}$ & $\boldsymbol{0.955}$ \\
    Choice 3 & $\boldsymbol{0.000}$ & 0.025 & $\boldsymbol{0.955}$ & Choice 3 &$\boldsymbol{0.000}$ & $\boldsymbol{0.025}$ & $\boldsymbol{0.955}$ \\
    Choice 4 & $\boldsymbol{0.000}$ & 0.026 & $\boldsymbol{0.951}$ & Choice 4 & $\boldsymbol{0.000}$ & $\boldsymbol{0.026}$ & $\boldsymbol{0.952}$ \\
    Choice 5 & $\boldsymbol{0.001}$ & $\boldsymbol{0.013}$ & $\boldsymbol{0.952}$ & Choice 5 & $\boldsymbol{-0.004}$ & $\boldsymbol{0.025}$ & $\boldsymbol{0.946}$ \\
    Choice 6 & $\boldsymbol{-0.001}$ & $\boldsymbol{0.013}$  & $\boldsymbol{0.953}$ & Choice 6 & $\boldsymbol{-0.001}$ & $\boldsymbol{0.025}$ & $\boldsymbol{0.950}$ \\
    Choice 7 & -0.041 & 0.028 & 0.943 & Choice 7 & $\boldsymbol{-0.001}$ & $\boldsymbol{0.026}$ & $\boldsymbol{0.951}$ \\
    Choice 8 & $\boldsymbol{0.001}$ & 0.025 & $\boldsymbol{0.948}$ & Choice 8 & $\boldsymbol{0.001}$ & $\boldsymbol{0.025}$ & $\boldsymbol{0.949}$ \\
    \hline $X_3$ & & &  & $X_7$ & & & \\
    \hline Choice 1 & 0.499 & 0.267 & 0.049 & Choice 1 & $\boldsymbol{0.002}$ & 0.046 & $\boldsymbol{0.953}$ \\
    Choice 2 & 0.500 & 0.267 & 0.050 & Choice 2 & $\boldsymbol{0.002}$ & 0.045 & $\boldsymbol{0.951}$ \\
    Choice 3 & $\boldsymbol{0.000}$ & $\boldsymbol{0.025}$ & $\boldsymbol{0.954}$ & Choice 3 &$\boldsymbol{0.000}$ & $\boldsymbol{0.025}$ & $\boldsymbol{0.954}$ \\
    Choice 4 & $\boldsymbol{0.000}$ & $\boldsymbol{0.026}$ & $\boldsymbol{0.951}$ & Choice 4 &$\boldsymbol{0.000}$ & $\boldsymbol{0.026}$ & $\boldsymbol{0.952}$  \\
    Choice 5 & 0.502 & 0.269 & 0.050 & Choice 5 & $\boldsymbol{0.002}$ & 0.046 & $\boldsymbol{0.951}$ \\
    Choice 6 & 0.499 & 0.267 & 0.051 & Choice 6 & $\boldsymbol{-0.002}$ & 0.046 & $\boldsymbol{0.949}$ \\
    Choice 7 & $\boldsymbol{-0.001}$ & $\boldsymbol{0.026}$ & $\boldsymbol{0.952}$ & Choice 7 & $\boldsymbol{-0.001}$ & $\boldsymbol{0.026}$& $\boldsymbol{0.951}$\\
    Choice 8 & $\boldsymbol{0.001}$ & $\boldsymbol{0.025}$ & $\boldsymbol{0.948}$ & Choice 8 & $\boldsymbol{0.001}$ & $\boldsymbol{0.025}$ & $\boldsymbol{0.949}$ \\
    \hline $X_4$ & & & & $X_8$ & & & \\
    \hline Choice 1 & $\boldsymbol{-0.001}$ & $\boldsymbol{0.013}$ & $\boldsymbol{0.949}$ &   Choice 1 & $\boldsymbol{-0.003}$ & $\boldsymbol{0.025}$ & $\boldsymbol{0.947}$ \\
    Choice 2 & $\boldsymbol{0.002}$ & $\boldsymbol{0.013}$ & $\boldsymbol{0.952}$ & Choice 2 & $\boldsymbol{0.001}$ & $\boldsymbol{0.024}$ & $\boldsymbol{0.955}$ \\
    Choice 3 &$\boldsymbol{0.000}$ & 0.025 & $\boldsymbol{0.954}$ & Choice 3 &$\boldsymbol{0.000}$ & $\boldsymbol{0.025}$ & $\boldsymbol{0.954}$ \\
    Choice 4 &$\boldsymbol{0.000}$ & 0.026 & $\boldsymbol{0.951}$ &  Choice 4 &$\boldsymbol{0.000}$ & $\boldsymbol{0.026}$ & $\boldsymbol{0.952}$ \\
    Choice 5 & $\boldsymbol{0.002}$ & $\boldsymbol{0.012}$ & $\boldsymbol{0.952}$ & Choice 5 & $\boldsymbol{0.000}$ &$\boldsymbol{0.025}$ & $\boldsymbol{0.954}$ \\
    Choice 6 & $\boldsymbol{-0.001}$ & $\boldsymbol{0.013}$ & $\boldsymbol{0.953}$ & Choice 6 & $\boldsymbol{-0.001}$ & $\boldsymbol{0.025}$ & $\boldsymbol{0.950}$ \\
    Choice 7 & $\boldsymbol{-0.001}$ & 0.026 & $\boldsymbol{0.952}$ & Choice 7 & $\boldsymbol{-0.001}$& $\boldsymbol{0.026}$ & $\boldsymbol{0.951}$ \\
    Choice 8 & $\boldsymbol{0.001}$ & 0.025 & $\boldsymbol{0.948}$ & Choice 8 & $\boldsymbol{0.001}$ & $\boldsymbol{0.025}$& $\boldsymbol{0.949}$ \\
    \hline \hline 
    \end{tabular}}
    \caption{Bias, MSE, and Coverage of $\widehat{\Delta}$. \textbf{Bolded} quantities indicate optimal values.}
     \label{table:individual_sim}
\end{table}
\pagebreak
\begin{table}[!htbp]
    \centering
    \scalebox{0.8}{
    \begin{tabular}{c|llllll}
    \hline & Full & Separate & Shared  & Sufficient  & Efficient & Null \\
    \hline \hline Bias &  -0.002 ($\pm 0.005$) & 0.025 ($\pm 0.004$) & 0.031 ($\pm 0.004$) & 0.126 ($\pm 0.006$) & 0.022 ($\pm 0.004$) & 0.407 ($\pm 0.004$)   \\
    MSE & 0.029 ($\pm 0.001$)  & 0.018 ($\pm 0.001$) & 0.024 ($\pm 0.001$) & 0.059 ($\pm 0.002$) & 0.017 ($\pm 0.001$) & 0.185 ($\pm 0.003$)   \\
    Coverage & 0.948  & 0.949 & 0.941  & 0.876 & 0.937  & 0.180 \\
    Predictors included ($\mu^{\text{diff}}$) & 8 & 4.12 ($\pm 0.01$) & 5.78 ($\pm 0.01$) & 1.69 ($\pm 0.01$) & 4.13 ($\pm 0.01$) & 0  \\
    Predictors included ($\mu$) & 8 & 4.15 ($\pm 0.01$) & 5.78 ($\pm 0.01$) & 0.52 ($\pm 0.01$) & 1.75 ($\pm 0.01$) & 0  \\
    \hline \hline
    \end{tabular}}
\caption{Bias, MSE, and coverage rates (with Margin of Error) of $\widehat{\Delta}$ using the model with all predictors (Full), each variable selection method, and the model with no predictors (Null), each with their mean number of predictors included, with $J=50$ groups}    
\label{table:study2_J50}
\end{table}

\begin{table}[!htbp]
    \centering
    \begin{tabular}{c | llllllll}
  \hline   &    $\bX_1$ & $\bX_2$ & $\bX_3$ & $\bX_4$ & $\bX_5$ & $\bX_6$ & $\bX_7$ & $\bX_8$  \\
 \hline \hline $\beta_k$ & 1 & 0 & 1 & 0 & 1 & 0 & 1 & 0 \\
    $\widetilde{\beta}_k$ & 1 & 1 & 0 & 0 & 1 & 1 & 0 & 0 \\
    $\alpha_k$ & 1 & 1 & 1  & 1 & 0 & 0 & 0 & 0 \\
   \hline Separate & 0.929 & 0.085 & 0.934 & 0.085 & 0.984 & 0.062 & 0.984 & 0.059  \\
   Shared   & 0.993 & 0.901 & 0.876 & 0.063 & 1.000 & 0.969 & 0.958 & 0.023 \\
Sufficient  & 0.695 & 0.139 & 0.681 & 0.141 & 0.012 & 0.003 & 0.012 & 0.003 \\
Efficient & 0.936 & 0.085 & 0.932 & 0.086 & 0.986 & 0.064 & 0.984 & 0.059 \\
\hline \hline 
   \end{tabular}
    \caption{Inclusion probabilities of $\bX_1, ..., \bX_8$ in the model for $\bmudiff$ using each method with corresponding true values of $\boldsymbol{\beta}$, $\boldsymbol{\widetilde{\beta}}, \boldsymbol{\alpha}$, with $J=50$ groups}
    \label{table:inclusion_mudiff_J50}
\end{table}

\begin{table}[!htbp]
    \centering
    \scalebox{0.8}{\begin{tabular}{c|llllll}
    \hline & Full & Separate & Shared  & Sufficient  & Efficient & Null \\
    \hline \hline Bias & 0.003 ($\pm 0.003$) & 0.003 ($\pm 0.002$) &  0.004 ($\pm 0.002$) & 0.008 ($\pm 0.003$) & 0.000 ($\pm 0.002$) & 0.410  ($\pm 0.003$)  \\
    MSE & 0.014 ($\pm 0.001$) & 0.005 ($\pm 0.000$) & 0.008 ($\pm 0.000$) & 0.011 ($\pm 0.001$) & 0.005 ($\pm 0.000$) & 0.177  ($\pm 0.002$) \\
    Coverage &  0.947 & 0.958 & 0.953 & 0.954 & 0.958  & 0.017   \\
    Predictors included ($\mu^{\text{diff}}$) & 8 & 4.18 ($\pm 0.00$)  & 6.02 ($\pm 0.00$)  & 2.14 ($\pm 0.00$)  & 4.18 ($\pm 0.00$)  & 0  \\
    Predictors included ($\mu$) & 8 & 4.17 ($\pm 0.00$)  & 6.02 ($\pm 0.00$)  & 1.04 ($\pm 0.01$)  & 2.20 ($\pm 0.01$)  & 0  \\
    \hline \hline
    \end{tabular}}
\caption{Bias, MSE, and coverage rates (with Margin of Error) of $\widehat{\Delta}$ using the model with all predictors (Full), each variable selection method, and the model with no predictors (Null), each with their mean number of predictors included, with $J=100$ groups}    
\label{table:study2_J100}
\end{table}

\begin{table}[!htbp]
    \centering
    \begin{tabular}{c | llllllll}
  \hline   &    $\bX_1$ & $\bX_2$ & $\bX_3$ & $\bX_4$ & $\bX_5$ & $\bX_6$ & $\bX_7$ & $\bX_8$  \\
 \hline \hline $\beta_k$ & 1 & 0 & 1 & 0 & 1 & 0 & 1 & 0 \\
    $\widetilde{\beta}_k$ & 1 & 1 & 0 & 0 & 1 & 1 & 0 & 0 \\
    $\alpha_k$ & 1 & 1 & 1  & 1 & 0 & 0 & 0 & 0 \\
   \hline Separate &  0.998 & 0.051 & 0.998 & 0.052 & 1.000 & 0.040 & 1.000 & 0.041  \\
   Shared   & 1.000 & 0.995 & 0.991 & 0.021 & 1.000 & 1.000 & 1.000 & 0.009  \\
Sufficient  & 0.975 & 0.071 & 0.970 & 0.071 & 0.025 & 0.001 & 0.023 & 0.001   \\
Efficient & 0.999 & 0.052 & 0.998 & 0.052 & 1.000 & 0.042 & 1.000 & 0.041 \\
\hline \hline 
   \end{tabular}
    \caption{Inclusion probabilities of $\bX_1, ..., \bX_8$ in the model for $\bmudiff$ using each method with corresponding true values of $\boldsymbol{\beta}$, $\boldsymbol{\widetilde{\beta}}, \boldsymbol{\alpha}$, with $J=100$ groups}
    \label{table:inclusion_mudiff_J100}
\end{table}
\pagebreak
\begin{table}[!htbp]
    \centering
    \begin{tabular}{cccccccccc}
       \hline & Age & Female & IVD & Type 1 & Commercial & Tobacco & Wealth & Income & $\boldsymbol{c}^{\textbf{(2008)}}$  \\
      \hline A1c & & & & & & & & & \\
      \hline Separate & 0.021 & 0.031 & 0.337 & 0.061 & 0.027 & 0.104 & 0.035 & 0.063 & 0.008 \\
      Shared & 0.001 & 0.978 & 0.156 & 0.004 & 0.001 & 0.030 & 0.003 & 0.003 & 0.000 \\
      Sufficient & 0.003 & 0.008 & 0.033 & 0.041 & 0.014 & 0.033 &  0.007 & 0.003 & 0.009 \\
      Efficient & 0.017 & 0.027 & 0.282 & 0.071 & 0.030 & 0.106 & 0.036 & 0.041  & 0.008 \\
      \hline LDL & & & & & & & & & \\
      \hline Separate & 0.430 & 0.128  &1.000& 0.324 &0.033 &0.139 &0.053& 0.119& 0.018 \\
      Shared & 0.529 & 0.956 & 1.000 & 0.337 & 0.003 & 0.050 & 0.001 & 0.002 & 0.001 \\
      Sufficient & 0.013 & 0.010 & 0.054 & 0.244 & 0.016 & 0.084 & 0.007 & 0.006 & 0.053 \\
      Efficient & 0.674 & 0.043 & 1.000 & 0.158 & 0.033 & 0.130 & 0.055 & 0.098 & 0.019 \\
      \hline SBP & & & & & & & & & \\
      \hline Separate & 0.145 & 0.044 & 0.561 & 0.092 & 0.037 & 0.115 & 0.023 & 0.046 & 0.010 \\
      Shared & 0.066 & 0.002 & 0.109 & 0.012 & 0.002 & 0.020 & 0.001 & 0.001 & 0.000 \\
      Sufficient & 0.017 & 0.008 & 0.046 & 0.053 & 0.020 & 0.039 & 0.006 & 0.004 & 0.011 \\
      Efficient & 0.127 & 0.039 & 0.579 & 0.092 & 0.038 & 0.101 & 0.027 & 0.044 & 0.010 \\
      \hline
    \end{tabular}
    \caption{Inclusion probabilities for each candidate variable in the change model}
    \label{table:inclusion}
\end{table}

\begin{table}[!htbp]
    \centering
    \scalebox{1}{
    \begin{tabular}{cccccc}
     \hline &  Separate & Shared & Sufficient & Efficient &  Full \\
    \hline \hline  $\boldsymbol{c}^{\textbf{diff}}$ ($\Delta$) & (-0.02, 0.03) & (-0.02, 0.03) & (-0.02, 0.03) & (-0.02, 0.03) & (-0.01, 0.05) \\
    Age & (-0.06, 0.04) & (-0.05, 0.04) & (-0.05, 0.04) & (-0.05, 0.04) & (-0.24, 0.11) \\
    Female &  (-0.05, 0.06) & (-0.16, 0.34) & (-0.05, 0.05) & (-0.06, 0.05) & (-0.24, 0.30) \\
    IVD & (-0.82, 0.04) & (-0.86, 0.04) & (-0.07, 0.043) & (-0.82, 0.04) & (-0.96, 0.01) \\
    Type 1 & (-0.05, 0.24) &  (-0.05, 0.05) & (-0.05, 0.10) &  (-0.05, 0.33) & -(0.18, 1.10) \\
    Commercial & (-0.06, 0.04) & (-0.05, 0.04) & (-0.06, 0.04) & (-0.07, 0.04) & (-0.25, 0.14) \\ 
    Tobacco & (-0.05, 0.56) & (-0.05, 0.62) & (-0.05, 0.07) & (-0.05, 0.53) & (-0.55, 0.85) \\
    Wealth & (-0.07, 0.04) & (-0.06, 0.04) & (-0.06, 0.04) & (-0.08, 0.04) & (-0.24, 0.12) \\
    Income & (-0.13, 0.03) & (-0.06, 0.03) & (-0.06, 0.03) & (-0.09, 0.03) & (-0.22, 0.06) \\
    $\boldsymbol{c}^{\textbf{(2008)}}$ & (-0.01, 0.05) & (-0.01, 0.04) & (-0.01, 0.05) & (-0.01, 0.05) & (-0.02, 0.05) \\
    \hline \hline 
    \end{tabular}}
    \caption{95$\%$ credible intervals for $\Delta$ and $\bBeta$ using each variable selection method, A1c}
    \label{table:CIs_A1c}
\end{table}
\pagebreak
\begin{table}[!htbp]
    \centering
    \scalebox{1}{
    \begin{tabular}{cccccc}
     \hline & Separate & Shared & Sufficient & Efficient & Full \\
    \hline \hline  $\boldsymbol{c}^{\textbf{diff}}$ ($\Delta$)  &  \textbf{(-0.07, -0.00)} & \textbf{(-0.07, -0.01)} & \textbf{(-0.09, -0.01)} & \textbf{(-0.07, -0.00)} & (-0.06, 0.02)  \\
    Age & (-0.03, 0.54) & (-0.01, 0.67) & (-0.04, 0.06) & (-0.02, 0.53) & \textbf{(0.25, 0.73)}  \\
    Female & (-0.04, 0.51) & (-0.04, 0.71) & (-0.05, 0.05) & (-0.05, 0.11) & \textbf{(0.01, 0.69)} \\
    IVD & \textbf{(-2.38, -1.04)} & \textbf{(-2.43, -1.16)} & (-1.57, 0.04) & \textbf{(-2.34, -1.06)} & \textbf{(-2.23, -0.93)} \\
    Type 1 & (-1.45, 0.05) & (-1.52, .05) &  (-1.52, 0.04) & (-1.06, 0.06)  & (-0.76, 1.03) \\
    Commercial & (-0.05, 0.06) & (-0.05, 0.05) & (-0.05, 0.05) & (-0.05, 0.06) & (-0.04, 0.61) \\ 
    Tobacco & (-0.05, 1.02) & (-0.05, 1.14) & (-0.05, 1.18) & (-0.06, 0.82) & (-0.04, 1.82) \\
    Wealth & (-0.17, 0.04) & (-0.06, 0.04) & (-0.06, 0.04) & (-0.17, 0.04) & (-0.32, 0.17) \\
    Income & (-0.25, 0.03) & (-0.06, 0.03) & (-0.06, 0.04) & (-0.22, 0.03) & (-0.33, 0.06) \\
    $\boldsymbol{c}^{\textbf{(2008)}}$ & (-0.01, 0.06) & (-0.01, 0.06) & (-0.00, 0.09) & (-0.01, 0.06) & \textbf{(0.00, 0.10)} \\
    \hline \hline 
    \end{tabular}}
    \caption{95$\%$ credible intervals for $\Delta$ and $\bBeta$ using each variable selection method, LDL}
    \label{table:CIs_LDL}
\end{table}

\begin{table}[!htbp]
    \centering
    \scalebox{1}{
    \begin{tabular}{cccccc}
     \hline & Separate & Shared & Sufficient & Efficient & Full \\
    \hline \hline  $\boldsymbol{c}^{\textbf{diff}}$ ($\Delta$)  & \textbf{(-0.09, -0.02)} & \textbf{(-0.09, -0.02)} & \textbf{(-0.09, -0.02)} & \textbf{(-0.09, -0.02)} & \textbf{(-0.11, -0.03)}\\
    Age & (-0.04, 0.35) & (-0.04, 0.34) & (-0.04, 0.06) & (-0.04, 0.34) &  \textbf{(0.07, 0.59)} \\
    Female & (-0.05, 0.10) & (-0.05, 0.05) & (-0.05, 0.05) & (-0.05, 0.12) & (-0.05, 0.68) \\
    IVD & (-0.04, 1.36) & (-0.04, 0.95) & (-0.04, 0.75) & (-0.04, 1.34) & \textbf{(0.04, 1.44)} \\
    Type 1 & (-0.50, 0.07) &  (-0.05, 0.05) & (-0.24, 0.06) & (-0.50, 0.07) & (-0.96, 0.99) \\ 
    Commercial & (-0.05, 0.09) & (-0.05, 0.05) & (-0.05, 0.06) & (-0.05, 0.08) & (-0.22, 0.49) \\ 
    Tobacco & (-0.68, 0.06) & (-0.08, 0.05) & (-0.38, 0.05) & (-0.700, 0.06) &  (-1.03, 0.94) \\
    Wealth & (-0.05, 0.06) & (-0.05, 0.05) & (-0.05, 0.05) & (-0.05, 0.06) & (-0.21, 0.31) \\
    Income & (-0.04, 0.13) &  (-0.04, 0.06) & (-0.04, 0.06) & (-0.04, 0.12) & (-0.11, 0.30) \\
    $\boldsymbol{c}^{\textbf{(2008)}}$ &  (-0.05, 0.02) & (-0.05, 0.02) & (-0.05, 0.02) & (-0.05, 0.02) & (-0.08,  0.02) \\
    \hline \hline 
    \end{tabular}}
    \caption{95$\%$ credible intervals for $\Delta$ and $\bBeta$ using each variable selection method, SBP}
    \label{table:CIs_SBP}
\end{table}

\appendix
\setcounter{table}{0}
\renewcommand{\thetable}{A\arabic{table}}
\clearpage
\section{Omitted variable bias: derivation and a simulation}\label{sec:appendixA}

Here, we expand on the technical details behind the omitted-variable bias result from Section~\ref{sec:omitted}. Recall our main result:

\begin{theorem*}
Suppose $\bY \sim \boldsymbol{N}(\boldsymbol{AB}\bTheta, \bSigma)$, where $\boldsymbol{B}$ is a matrix and $\bTheta$ is a vector, each which can be partitioned into $[\boldsymbol{B_1} \ \boldsymbol{B_0}]$ and $[\bTheta_1 \ \bTheta_0]^T$ respectively. Suppose $\bY$ is modeled as $\bY \sim \boldsymbol{N}(\boldsymbol{AB_1}\bTheta_1, \bSigma)$. Finally, let $\bTheta_1$ have a flat prior. Then, the bias of $\widehat{\bTheta}_1$ as estimated by the model is \\
\begin{equation*}
\mathbb{E}_{\bY \vert \bTheta}[\widehat{\bTheta}_1 - \bTheta_1] = 
(\boldsymbol{B_1}^T \boldsymbol{A}^T \boldsymbol{\Sigma}^{-1} \boldsymbol{AB_1})^{-1}\boldsymbol{B_1}^T \boldsymbol{A}^T \boldsymbol{\Sigma}^{-1} \boldsymbol{AB_0} \bTheta_0.
\end{equation*}
\end{theorem*}
The proof of the result is straightforward. The posterior distribution of $\widehat{\bTheta}_1$ is 
\begin{equation*}
    p(\widehat{\bTheta}_1 \vert \bY) = \boldsymbol{N}(\boldsymbol{V}\boldsymbol{B_1}^T\boldsymbol{A}^T\bSigma^{-1}\bY, \boldsymbol{V}), \mbox{ where } \boldsymbol{V} \equiv (\boldsymbol{B_1}^T\boldsymbol{A}^T\bSigma^{-1}\boldsymbol{AB_1})^{-1}
\end{equation*}
Then, $\mathbb{E}_{\bY \vert \bTheta}[\widehat{\bTheta}_1 - \bTheta_1] $ \\
$= \boldsymbol{V}\boldsymbol{B_1}^T\boldsymbol{A}^T\bSigma^{-1}\mathbb{E}_{\bY \vert \bTheta}\bY - \bTheta_1$ \\
$= \boldsymbol{V}\boldsymbol{B_1}^T\boldsymbol{A}^T\bSigma^{-1} \boldsymbol{A} (\boldsymbol{B_1}\bTheta_1 + \boldsymbol{B_0} \bTheta_0) - \bTheta_1 $  \\
$= (\boldsymbol{B_1}^T \boldsymbol{A}^T \boldsymbol{\Sigma}^{-1} \boldsymbol{AB_1})^{-1}\boldsymbol{B_1}^T \boldsymbol{A}^T \boldsymbol{\Sigma}^{-1} \boldsymbol{AB_0} \bTheta_0 \ \blacksquare$ \\

In our context, $\bTheta_1 \equiv [\bBetatilde_{\boldsymbol{\widetilde{w}} = 1} \ \widetilde{\Delta}_1 \ \bBeta_{\boldsymbol{w} = 1} \ \Delta]^T]$ and $\bTheta_0 \equiv [\bBetatilde_{\boldsymbol{\widetilde{w}} = 0} \ \widetilde{\Delta}_0 \ \bBeta_{\boldsymbol{w} = 0}]^T$ where $\widetilde{\Delta}_1$ = $\widetilde{\Delta}$ when $\bT$ is included in the baseline model and 0 otherwise and $\widetilde{\Delta}_0$ = $\widetilde{\Delta}$ when $\bT$ is excluded from the baseline model and 0 otherwise, 
$\boldsymbol{Y} \equiv [ Y_{11}^{(0)}, ..., Y_{J,n_J^{(0)}}^{(0)},Y_{11}^{(1)}, ..., Y_{J,n_J^{(1)}}^{(1)}]^T$, $\boldsymbol{A} \equiv 
\begin{bmatrix}
\boldsymbol{A_0} & \boldsymbol{0} \\
\boldsymbol{0} & \boldsymbol{A_1} 
\end{bmatrix} $ where $\boldsymbol{A_0}$ and $\boldsymbol{A_1}$ are matrices that assign pre- and post-treatment group-level predictors to the subjects with that group (respectively), $\boldsymbol{B_1} \equiv
\begin{bmatrix}
\bX_{\boldsymbol{\widetilde{w}} = 1} & \widetilde{\bT} & \boldsymbol{0} & \boldsymbol{0} \\
\bX_{\boldsymbol{\widetilde{w}} = 1} & \widetilde{\bT} & \boldsymbol{X}_{\boldsymbol{w} = 1} & \boldsymbol{T} 
\end{bmatrix}$ such that $\bX_{\boldsymbol{\widetilde{w}} = 1}$ and $\boldsymbol{X}_{\boldsymbol{w} = 1}$ are those covariates that are included in the baseline and change models (respectively) and $\widetilde{\bT} = \bT$ when $\bT$ is included in the baseline model and $\boldsymbol{0}$ when it is not, $\boldsymbol{B_0} \equiv
\begin{bmatrix}
\boldsymbol{\widetilde{U}} & \boldsymbol{0} \\
\boldsymbol{\widetilde{U}} & \boldsymbol{U} \\
\end{bmatrix}$ where $\boldsymbol{\widetilde{U}}$ and $\boldsymbol{U}$ are the covariates excluded from the baseline and change models, respectively, 
 and $\boldsymbol{\Sigma} \equiv $
$
\begin{bmatrix} 
\boldsymbol{\Sigma}_{11} & \boldsymbol{\Sigma}_{12} \\
\boldsymbol{\Sigma}_{21} & \boldsymbol{\Sigma}_{22}
\end{bmatrix}
$, where $\boldsymbol{\Sigma}_{11}$ is a block diagonal matrix where the j'th block is a diagonal matrix with $\widetilde{\sigma}_j + \widetilde{\tau}^2$ on the diagonal and $\widetilde{\tau}^2$ on the within-block off-diagonals, $\boldsymbol{\Sigma}_{12}$  and $\boldsymbol{\Sigma}_{21}$ are block diagonal matrices with every element in each block being $\widetilde{\tau}^2$, and $\boldsymbol{\Sigma}_{22}$ is a block diagonal matrix where the j'th block is a diagonal matrix with $\sigma_j^2 + \tau^2 + \widetilde{\tau}^2$ on the diagonal and $\tau^2 + \widetilde{\tau}^2$ on the within-block off-diagonals.  

Isolating $\Delta$ is difficult, so we perform a brief simulation to see which covariates we need to include when we sequentially add predictors to both $\bmu$ and $\bmudiff$, comparing the bias of $\widehat{\Delta}$ when we (1) do not adjust the baseline model for $\bT$, and (2) adjust the baseline model for $\bT$. The simulations proceed by, for each iteration, generating data via Section~\ref{sec:data_generation} and computing the bias result above. 

\begin{table}[ht]
\centering
    \begin{tabular}{c|cc}
      \hline  & No adjust for $\bT$ & Adjust for $\bT$  \\
      \hline \hline Null & 0.437 & 0.400  \\
        + $X_1$ & 0.271 & 0.250  \\
        + $X_3$ & 0.031 & -0.001 \\
        + $X_2$ & 0.002 & 0.000 \\
        + $X_5$ & 0.000 & 0.002 \\
        + $X_7$ & 0.000 & 0.000  \\
        + $X_6$ & 0.000 & 0.000 \\
        + $X_4$ & 0.000 & 0.000  \\
        Full & 0.000 & 0.000  \\
        \hline \hline 
    \end{tabular}
    \caption{Bias of $\widehat{\Delta}$ by including the current and all preceding rows as covariates for $\bmu$ and $\bmudiff$, without and with adjusting $\bmu$ for $\bT$}
    \label{table:baseline_sim_50}
\end{table}

Table~\ref{table:baseline_sim_50} shows the results of this simulation, where each row correponds to the addition of a new covariate into the baseline and change models and the columns correspond to whether or not we adjust $\bmu$ for $\bT$. As an example, the ``+ $X_3$'' and ``Adjust for $\bT$'' cell adjusts the model for $\bmu$ and $\bmudiff$ with $\{\bT, \bX_1, \bX_3 \}$ as covariates. The results show, and consistent with Table~\ref{table:individual_sim}, that when we do not adjust $\bmu$ for $\bT$, we must adjust $\bmu$ for all covariates predictive of both $\bmu$ and $\bT$ \textit{and} adjust $\bmudiff$ for all covariates predictive of both $\bmudiff$ and $\bT$ (here, $\{\bX_1, \bX_2, \bX_3 \}$) to achieve unbiasedness in estimating $\Delta$. When we do adjust $\bmu$ for $\bT$, Tables~\ref{table:individual_sim} and \ref{table:baseline_sim_50} suggest that we only need to adjust for all covariates predictive of $\bT$ and $\bmudiff$ (here, $\{\bX_1, \bX_3 \}$) to acheive unbiasedness. 

\newpage
\section{Gibbs algorithms for each variable selection method}\label{sec:appendixB}
This appendix outlines the Gibbs sampling algorithms used in the simulation studies outlined in Section~\ref{sec:simulations}. 
\subsection*{Separate method}
Initialize. \\ 
For iterations $t = 2, ..., T$, \\
For clinics $j = 1, ..., J$, \\
\indent Draw $\mu_j^{(t)} \sim N(\frac{\widetilde{\tau}^{2, (t-1)} \{\sigma_j^{2, (t-1)} n_j^{( 0)}\bar{\boldsymbol{Y}}_{j.}^{( 0)}+ \widetilde{\sigma}_j^{2, (t-1)} n_j^{(1)}(\bar{\boldsymbol{Y}}_{j.}^{(1)}- \mu_j^{\text{diff},(t-1)}) \} + \widetilde{\sigma}_j^{2, (t-1)} \sigma_j^{2, (t-1)} \bX_j \bBetatilde^{(t-1)}}{\widetilde{\tau}^{2, (t-1)}(n_j^{( 0)}\sigma_j^{2, (t-1)} + n_j^{(1)}\widetilde{\sigma}^{2, (t-1)}) + \widetilde{\sigma}_j^{2, (t-1)} \sigma_j^{2, (t-1)}}, \\ 
\hfill \frac{\widetilde{\sigma}_j^{2, (t-1)} \sigma_j^{2, (t-1)} \wildetilde{\tau}^{2, (t-1)}}{\widetilde{\tau}^{2, (t-1)} (n_j^{( 0)}\sigma_j^{2, (t-1)} + n_j^{(1)}\widetilde{\sigma}_j^{2, (t-1)}) + \widetilde{\sigma}_j^{2, (t-1)} \sigma_j^{2, (t-1)}})$ \\
\indent Draw $\mu_j^{\text{diff},(t)} \sim N(\frac{n_j^{(1)}\tau^{2, (t-1)} (\bar{\boldsymbol{Y}}_{j.}^{(1)}- \mu_j^{(t)}) + \sigma_j^{2, (t-1)} \bX_j \bBeta}{n_j^{(1)}\tau^{2, (t-1)} + \sigma_j^{2, (t-1)}}, \frac{\sigma_j^{2, (t-1)} \tau^{2, (t-1)}}{n_j^{(1)}\tau^{2, (t-1)} + \sigma_j^{2, (t-1)}})$ \\
\indent Draw $\widetilde{\sigma}_j^{2, (t)} \sim IG(0.5 n_j^{(0)}, 0.5 \sum_{i=1}^{n_j^{(0)}} (Y_{ji}^{( 0)}- \mu_j^{(t)})^2)$ \\ 
\indent Draw $\sigma_j^{2, (t)} \sim IG(0.5 n_j^{(1)}, 0.5 \sum_{i=1}^{n_j^{(1)}} (Y_{ji}^{(1)}- \mu_j^{(t)} - \mu_j^{\text{diff},(t)})^2)$ \\ 
Draw $\widetilde{\tau}^{2, (t)} \sim IG(0.5J, 0.5\vert \vert \bmu^{(t)} - \bX \bBetatilde^{(t-1)} \vert \vert _2$) \\ 
Draw $\tau^{2, (t)} \sim IG(0.5J, 0.5\vert \vert \bmu^{\text{diff}, (t)} - \bX \bBeta^{(t-1)}\vert \vert _2$) \\ 
Draw $\bBetatilde^{(t)} \sim \boldsymbol{N} ( \{ \widetilde{\tau}^{-2, (t)} \bX ^T \bX + \boldsymbol{\widetilde{D}}^{-2} \}^{-1} \widetilde{\tau}^{-2, (t)} \bX ^T \bmu^{(t)}, \{ \widetilde{\tau}^{-2, (t)} \bX ^T \bX + \boldsymbol{\widetilde{D}}^{-2} \}^{-1})$ \\
Draw $\bBeta^{(t)}\sim \boldsymbol{N} ( \{ \tau^{-2, (t)} \bX ^T \bX + \boldsymbol{D}^{-2} \}^{-1} \tau^{-2, (t)} \bX ^T \bmu^{\text{diff}, (t)}, \{ \tau^{-2, (t)} \bX ^T \bX + \boldsymbol{D}^{-2} \}^{-1})$ \\
For candidate variables k = 1, ... K, \\
\indent Draw $\widetilde{w}_k^{(t)}\sim \text{Bern} (\frac{1}{1 + \widetilde{BF}_k})$, with $\widetilde{BF}_k = \frac{N(\widetilde{\beta}_k^{(t)} \vert 0, 0.1^2)}{N(\widetilde{\beta}_k^{(t)} \vert 0, 1/\widetilde{\gamma}_k^{(t)})}$ \\
\indent Draw $w_k^{(t)} \sim \text{Bern} (\frac{1}{1 + BF_k})$, where $BF_k = \frac{N(\beta_k^{(t)} \vert 0, 0.1^2)}{N(\beta_k^{(t)} \vert 0, 1/\gamma_k^{(t)})}$ \\
\indent Update $\boldsymbol{\widetilde{D}}$ and $\boldsymbol{D}$ based on the new draws for $\boldsymbol{\widetilde{w}}$ and $\boldsymbol{w}$. \\
\indent Draw $\widetilde{\gamma}_k^{2, (t)} \sim IG(2.5 + 0.5\widetilde{w}_k^{(t)}, 2.5*5^2 + 0.5 \widetilde{w}_k^{(t)} \widetilde{\beta}^{2,(t)}_k)$ \\
\indent Draw $\gamma_k^{2, (t)} \sim IG(2.5 + 0.5 w_k^{(t)}, 2.5*5^2 + 0.5 w_k^{(t)} \beta_k^{2, (t)})$ 
\subsection*{Shared method}
Initialize. \\ 
For iterations $t = 2, ..., T$, \\
For clinics $j = 1, ..., J$, \\
\indent Draw $\mu_j^{(t)} \sim N(\frac{\widetilde{\tau}^{2, (t-1)} \{\sigma_j^{2, (t-1)} n_j^{( 0)}\bar{\boldsymbol{Y}}_{j.}^{( 0)}+ \widetilde{\sigma}_j^{2, (t-1)} n_j^{(1)}(\bar{\boldsymbol{Y}}_{j.}^{(1)}- \mu_j^{\text{diff},(t-1)}) \} + \widetilde{\sigma}_j^{2, (t-1)} \sigma_j^{2, (t-1)} \bX_j \bBetatilde^{(t-1)}}{\widetilde{\tau}^{2, (t-1)}(n_j^{( 0)}\sigma_j^{2, (t-1)} + n_j^{(1)}\widetilde{\sigma}^{2, (t-1)}) + \widetilde{\sigma}_j^{2, (t-1)} \sigma_j^{2, (t-1)}}, \\ 
\hfill \frac{\widetilde{\sigma}_j^{2, (t-1)} \sigma_j^{2, (t-1)} \wildetilde{\tau}^{2, (t-1)}}{\widetilde{\tau}^{2, (t-1)} (n_j^{( 0)}\sigma_j^{2, (t-1)} + n_j^{(1)}\widetilde{\sigma}_j^{2, (t-1)}) + \widetilde{\sigma}_j^{2, (t-1)} \sigma_j^{2, (t-1)}})$ \\
\indent Draw $\mu_j^{\text{diff},(t)} \sim N(\frac{n_j^{(1)}\tau^{2, (t-1)} (\bar{\boldsymbol{Y}}_{j.}^{(1)}- \mu_j^{(t)}) + \sigma_j^{2, (t-1)} \bX_j \bBeta}{n_j^{(1)}\tau^{2, (t-1)} + \sigma_j^{2, (t-1)}}, \frac{\sigma_j^{2, (t-1)} \tau^{2, (t-1)}}{n_j^{(1)}\tau^{2, (t-1)} + \sigma_j^{2, (t-1)}})$ \\
\indent Draw $\widetilde{\sigma}_j^{2, (t)} \sim IG(0.5 n_j^{(0)}, 0.5 \sum_{i=1}^{n_j^{(0)}} (Y_{ji}^{( 0)}- \mu_j^{(t)})^2)$ \\ 
\indent Draw $\sigma_j^{2, (t)} \sim IG(0.5 n_j^{(1)}, 0.5 \sum_{i=1}^{n_j^{(1)}} (Y_{ji}^{(1)}- \mu_j^{(t)} - \mu_j^{\text{diff},(t)})^2)$ \\ 
Draw $\widetilde{\tau}^{2, (t)} \sim IG(0.5J, 0.5\vert \vert \bmu^{(t)} - \bX \bBetatilde^{(t-1)} \vert \vert _2$) \\ 
Draw $\tau^{2, (t)} \sim IG(0.5J, 0.5\vert \vert \bmu^{\text{diff}, (t)} - \bX \bBeta^{(t-1)}\vert \vert _2$) \\ 
Draw $\bBetatilde^{(t)} \sim \boldsymbol{N} ( \{ \widetilde{\tau}^{-2, (t)} \bX^T \bX + \boldsymbol{D}^{-2} \} ^{-1} \widetilde{\tau}^{-2, (t)} \bX ^T \bmu^{(t)}, \{ \widetilde{\tau}^{-2, (t)} \bX ^T \bX + \boldsymbol{D}^{-2} \}^{-1})$ \\
Draw $\bBeta^{(t)}\sim \boldsymbol{N} ( \{ \tau^{-2, (t)} \bX ^T \bX + \boldsymbol{D}^{-2} \}^{-1} \tau^{-2, (t)} \bX ^T \bmu^{\text{diff}, (t)}, \{ \tau^{-2, (t)} \bX ^T \bX + \boldsymbol{D}^{-2} \}^{-1})$ \\
For candidate variables k=1, ..., K, \\
\indent Draw $w_k^{(t)} \sim \text{Bern} (\frac{1}{1 + BF_k})$, with $BF_k = \frac{\boldsymbol{N_2}([\widetilde{\beta}^{(t)}_k, \beta_k^{(t)}]^T \vert \boldsymbol{0_2}, 0.1^2 \boldsymbol{I_2})}{\boldsymbol{N_2}([\widetilde{\beta}^{(t)}_k, \beta_k^{(t)}]^T \vert \boldsymbol{0_2}, 1/\widetilde{\gamma}_k^{(t)} \boldsymbol{I_2})}$ \\
\indent Update $\boldsymbol{D}$ based on the new draw for $\boldsymbol{w}$. \\
\indent Draw $\gamma_k^{2, (t)} \sim IG(2.5 + 0.5 w_k^{(t)}, 2.5*5^2 + 0.5 w_k^{(t)}(\widetilde{\beta}^{2,(t)}_k + \beta_k^{2, (t)}))$ 

\subsection*{Sufficient method}
Initialize. \\ 
For iterations $t = 2, ..., T$, \\
For clinics $j = 1, ..., J$, \\
\indent Draw $\mu_j^{(t)} \sim N(\frac{\widetilde{\tau}^{2, (t-1)} \{\sigma_j^{2, (t-1)} n_j^{( 0)}\bar{\boldsymbol{Y}}_{j.}^{( 0)}+ \widetilde{\sigma}_j^{2, (t-1)} n_j^{(1)}(\bar{\boldsymbol{Y}}_{j.}^{(1)}- \mu_j^{\text{diff},(t-1)}) \} + \widetilde{\sigma}_j^{2, (t-1)} \sigma_j^{2, (t-1)} \bX_j \bBetatilde^{(t-1)}}{\widetilde{\tau}^{2, (t-1)}(n_j^{( 0)}\sigma_j^{2, (t-1)} + n_j^{(1)}\widetilde{\sigma}^{2, (t-1)}) + \widetilde{\sigma}_j^{2, (t-1)} \sigma_j^{2, (t-1)}}, \\ 
\hfill \frac{\widetilde{\sigma}_j^{2, (t-1)} \sigma_j^{2, (t-1)} \wildetilde{\tau}^{2, (t-1)}}{\widetilde{\tau}^{2, (t-1)} (n_j^{( 0)}\sigma_j^{2, (t-1)} + n_j^{(1)}\widetilde{\sigma}_j^{2, (t-1)}) + \widetilde{\sigma}_j^{2, (t-1)} \sigma_j^{2, (t-1)}})$ \\
\indent Draw $\mu_j^{\text{diff},(t)} \sim N(\frac{n_j^{(1)}\tau^{2, (t-1)} (\bar{\boldsymbol{Y}}_{j.}^{(1)}- \mu_j^{(t)}) + \sigma_j^{2, (t-1)} \bX_j \bBeta}{n_j^{(1)}\tau^{2, (t-1)} + \sigma_j^{2, (t-1)}}, \frac{\sigma_j^{2, (t-1)} \tau^{2, (t-1)}}{n_j^{(1)}\tau^{2, (t-1)} + \sigma_j^{2, (t-1)}})$ \\
\indent Draw $\widetilde{\sigma}_j^{2, (t)} \sim IG(0.5 n_j^{(0)}, 0.5 \sum_{i=1}^{n_j^{(0)}} (Y_{ji}^{( 0)}- \mu_j^{(t)})^2)$ \\ 
\indent Draw $\sigma_j^{2, (t)} \sim IG(0.5 n_j^{(1)}, 0.5 \sum_{i=1}^{n_j^{(1)}} (Y_{ji}^{(1)}- \mu_j^{(t)} - \mu_j^{\text{diff},(t)})^2)$ \\ 
Draw $\widetilde{\tau}^{2, (t)} \sim IG(0.5J, 0.5\vert \vert \bmu^{(t)} - \bX \bBetatilde^{(t-1)} \vert \vert _2$) \\ 
Draw $\tau^{2, (t)} \sim IG(0.5J, 0.5\vert \vert \bmu^{\text{diff}, (t)} - \bX \bBeta^{(t-1)}\vert \vert _2$) \\ 
Draw $\bBetatilde^{(t)} \sim \boldsymbol{N} ( \{ \widetilde{\tau}^{-2, (t)} \bX ^T \bX + \boldsymbol{\widetilde{D}}^{-2} \}^{-1} \widetilde{\tau}^{-2, (t)} \bX ^T \bmu^{(t)}, \{ \widetilde{\tau}^{-2, (t)} \bX ^T \bX + \boldsymbol{\widetilde{D}}^{-2} \}^{-1})$ \\
Draw $\bBeta^{(t)}\sim \boldsymbol{N} ( \{ \tau^{-2, (t)} \bX ^T \bX + \boldsymbol{D}^{-2} \}^{-1} \tau^{-2, (t)} \bX ^T \bmu^{\text{diff}, (t)}, \{ \tau^{-2, (t)} \bX ^T \bX + \boldsymbol{D}^{-2} \}^{-1})$ \\
For candidate variables k = 1, ... K, \\
\indent Draw $w_k^e \sim \text{Bern} \bigg(\frac{1}{1 + BF_k^e} \bigg)$, where $BF_k^e = \frac{N(\alpha_k \vert 0, 0.1^2)}{N(\alpha_k \vert 0, 1/\gamma_k^e)}$ \\
\indent  Draw $w_k \sim w_k^e * \text{Bern} \bigg(\frac{1}{1 + BF_k} \bigg)$, where $BF_k = \frac{N(\beta_k \vert 0, 0.1^2)}{N(\beta_k \vert 0, 1/\gamma_k)}$ \\
\indent Draw $\widetilde{w}_k \sim w_k * \text{Bern} \bigg(\frac{1}{1 + \widetilde{BF}_k} \bigg)$, where $\widetilde{BF}_k = \frac{N(\widetilde{\beta}_k \vert 0, 0.1^2)}{N(\widetilde{\beta}_k \vert 0, 1/\widetilde{\gamma}_k)}$ \\
\indent Update $\boldsymbol{\widetilde{D}}$ and $\boldsymbol{D}$ based on the new draws for $\boldsymbol{\widetilde{w}}$ and $\boldsymbol{w}$. \\
\indent Draw $\widetilde{\gamma}_k^{2, (t)} \sim IG(2.5 + 0.5\widetilde{w}_k^{(t)}, 2.5*5^2 + 0.5 \widetilde{w}_k^{(t)} \widetilde{\beta}^{2,(t)}_k)$ \\
\indent Draw $\gamma_k^{2, (t)} \sim IG(2.5 + 0.5 w_k^{(t)}, 2.5*5^2 + 0.5 w_k^{(t)} \beta_k^{2, (t)})$ 

\subsection*{Efficient method}
Initialize. \\ 
For iterations $t = 2, ..., T$, \\
For clinics $j = 1, ..., J$, \\
\indent Draw $\mu_j^{(t)} \sim N(\frac{\widetilde{\tau}^{2, (t-1)} \{\sigma_j^{2, (t-1)} n_j^{( 0)}\bar{\boldsymbol{Y}}_{j.}^{( 0)}+ \widetilde{\sigma}_j^{2, (t-1)} n_j^{(1)}(\bar{\boldsymbol{Y}}_{j.}^{(1)}- \mu_j^{\text{diff},(t-1)}) \} + \widetilde{\sigma}_j^{2, (t-1)} \sigma_j^{2, (t-1)} \bX_j \bBetatilde^{(t-1)}}{\widetilde{\tau}^{2, (t-1)}(n_j^{( 0)}\sigma_j^{2, (t-1)} + n_j^{(1)}\widetilde{\sigma}^{2, (t-1)}) + \widetilde{\sigma}_j^{2, (t-1)} \sigma_j^{2, (t-1)}}, \\ 
\hfill \frac{\widetilde{\sigma}_j^{2, (t-1)} \sigma_j^{2, (t-1)} \wildetilde{\tau}^{2, (t-1)}}{\widetilde{\tau}^{2, (t-1)} (n_j^{( 0)}\sigma_j^{2, (t-1)} + n_j^{(1)}\widetilde{\sigma}_j^{2, (t-1)}) + \widetilde{\sigma}_j^{2, (t-1)} \sigma_j^{2, (t-1)}})$ \\
\indent Draw $\mu_j^{\text{diff},(t)} \sim N(\frac{n_j^{(1)}\tau^{2, (t-1)} (\bar{\boldsymbol{Y}}_{j.}^{(1)}- \mu_j^{(t)}) + \sigma_j^{2, (t-1)} \bX_j \bBeta}{n_j^{(1)}\tau^{2, (t-1)} + \sigma_j^{2, (t-1)}}, \frac{\sigma_j^{2, (t-1)} \tau^{2, (t-1)}}{n_j^{(1)}\tau^{2, (t-1)} + \sigma_j^{2, (t-1)}})$ \\
\indent Draw $\widetilde{\sigma}_j^{2, (t)} \sim IG(0.5 n_j^{(0)}, 0.5 \sum_{i=1}^{n_j^{(0)}} (Y_{ji}^{( 0)}- \mu_j^{(t)})^2)$ \\ 
\indent Draw $\sigma_j^{2, (t)} \sim IG(0.5 n_j^{(1)}, 0.5 \sum_{i=1}^{n_j^{(1)}} (Y_{ji}^{(1)}- \mu_j^{(t)} - \mu_j^{\text{diff},(t)})^2)$ \\ 
Draw $\widetilde{\tau}^{2, (t)} \sim IG(0.5J, 0.5\vert \vert \bmu^{(t)} - \bX \bBetatilde^{(t-1)} \vert \vert _2$) \\ 
Draw $\tau^{2, (t)} \sim IG(0.5J, 0.5\vert \vert \bmu^{\text{diff}, (t)} - \bX \bBeta^{(t-1)}\vert \vert _2$) \\ 
Draw $\bBetatilde^{(t)} \sim \boldsymbol{N} ( \{ \widetilde{\tau}^{-2, (t)} \bX ^T \bX + \boldsymbol{\widetilde{D}}^{-2} \}^{-1} \widetilde{\tau}^{-2, (t)} \bX ^T \bmu^{(t)}, \{ \widetilde{\tau}^{-2, (t)} \bX ^T \bX + \boldsymbol{\widetilde{D}}^{-2} \}^{-1})$ \\
Draw $\bBeta^{(t)}\sim \boldsymbol{N} ( \{ \tau^{-2, (t)} \bX ^T \bX + \boldsymbol{D}^{-2} \}^{-1} \tau^{-2, (t)} \bX ^T \bmu^{\text{diff}, (t)}, \{ \tau^{-2, (t)} \bX ^T \bX + \boldsymbol{D}^{-2} \}^{-1})$ \\
For candidate variables k = 1, ... K, \\
\indent Draw $w_k \sim \text{Bern} \bigg(\frac{1}{1 + BF_k} \bigg)$, where $BF_k = \frac{N(\beta_k \vert 0, 0.1^2)}{N(\beta_k \vert 0, 1/\gamma_k)}$ \\
\indent Draw $\widetilde{w}_k \sim w_k * \text{Bern} \bigg(\frac{1}{1 + \widetilde{BF}_k} \bigg)$, where $\widetilde{BF}_k = \frac{N(\widetilde{\beta}_k \vert 0, 0.1^2)}{N(\widetilde{\beta}_k \vert 0, 1/\widetilde{\gamma}_k)}$ \\
\indent Update $\boldsymbol{\widetilde{D}}$ and $\boldsymbol{D}$ based on the new draws for $\boldsymbol{\widetilde{w}}$ and $\boldsymbol{w}$. \\
\indent Draw $\widetilde{\gamma}_k^{2, (t)} \sim IG(2.5 + 0.5\widetilde{w}_k^{(t)}, 2.5*5^2 + 0.5 \widetilde{w}_k^{(t)} \widetilde{\beta}^{2,(t)}_k)$ \\
\indent Draw $\gamma_k^{2, (t)} \sim IG(2.5 + 0.5 w_k^{(t)}, 2.5*5^2 + 0.5 w_k^{(t)} \beta_k^{2, (t)})$

\end{document}